\documentclass[aps,prb,superscriptaddress,amsfonts,amsmath,amssymb,showpacs,floatfix,reprint,longbibliography,footinbib]{revtex4-2}

\usepackage{times}
\usepackage{url}
\usepackage{bm}
\usepackage{graphicx}
\usepackage{amsmath}
\usepackage{amstext}
\usepackage{amssymb}
\usepackage{amsfonts}
\usepackage{amsbsy}
\usepackage{verbatim}
\usepackage{xcolor}
\definecolor{dkgreen}{rgb}{0, 0.5, 0}
\definecolor{midnightblue}{rgb}{0.39, 0.58, 0.93}
\definecolor{darkblue}{HTML}{004D6B}
\definecolor{darkred}{HTML}{8c1515}
\definecolor{darkgreen}{HTML}{006400}
\usepackage[colorlinks=true, urlcolor=midnightblue, linkcolor=midnightblue, citecolor=midnightblue, pdftex]{hyperref}
\usepackage[capitalise]{cleveref}
\usepackage{multirow}
\usepackage{floatrow}
\usepackage{float}
\usepackage{tikz}
\usepackage{gensymb}
\usepackage{textcomp}
\usepackage{enumitem}
\usepackage[version=3]{mhchem}
\usepackage{afterpage}
\usepackage{pbox}
\usepackage{makecell}
\usepackage{dsfont}
\usepackage{siunitx}
\usepackage{stackengine,wasysym,scalerel}
\usepackage{latexsym}

\usepackage{cancel}
\usepackage{array, multirow, bigdelim, makecell, booktabs}
\usepackage[misc]{ifsym}
\usepackage[percent]{overpic}
\usepackage[normalem]{ulem}
\usepackage{placeins}

\usepackage{MnSymbol}

\graphicspath{{figures/}}

\begin{document}

\title{Quantum effects on pyrochlore higher-rank U(1) spin liquids:\\ 
	Pinch-line singularities, spin nematics, and connections to oxide materials}
  
\author{Lasse Gresista}
\email{gresista@thp.uni-koeln.de}
\affiliation{Institute for Theoretical Physics, University of Cologne, 50937 Cologne, Germany}
\affiliation{Department of Physics and Quantum Center for Diamond and Emergent Materials (QuCenDiEM), Indian Institute of Technology Madras, Chennai 600036, India}
\author{Daniel Lozano-G\'omez}
\affiliation{Institut f\"ur Theoretische Physik and W\"urzburg-Dresden Cluster of Excellence ct.qmat, Technische Universit\"at Dresden, 01062 Dresden, Germany}
\author{Matthias Vojta}
\affiliation{Institut f\"ur Theoretische Physik and W\"urzburg-Dresden Cluster of Excellence ct.qmat, Technische Universit\"at Dresden, 01062 Dresden, Germany}
\author{Simon Trebst}
\affiliation{Institute for Theoretical Physics, University of Cologne, 50937 Cologne, Germany}
\author{Yasir Iqbal}
\affiliation{Department of Physics and Quantum Center for Diamond and Emergent Materials (QuCenDiEM), Indian Institute of Technology Madras, Chennai 600036, India}

 \date{\today}

\begin{abstract}
Motivated by the magnetism of pyrochlore oxides, we consider the effect of quantum fluctuations in the most general symmetry-allowed nearest-neighbor Kramers exchange Hamiltonian on the pyrochlore lattice. At the classical level, this Hamiltonian exhibits a rich landscape of classical spin liquids and a variety of nonconventional magnetic phases. In contrast, much remains unclear for the quantum model, where quantum fluctuations have the potential to alter the classical landscape and stabilize novel magnetic phases. Employing state-of-the-art pseudo-fermion functional renormalization group calculations for the spin-$1/2$ model, we determine the quantum phase diagram at relevant cross-sections, where the classical model hosts an algebraic nodal rank-2 spin liquid and a spin nematic order. We find large regions in parameter space on which dipolar magnetic order is absent and, based on known fingerprints in the correlation functions, we suggest that this nonconventional region is composed of an ensemble of distinct phases stabilized by quantum fluctuations. Our results hint at the existence of a spin nematic phase, and we identify the quantum analog of the classical rank-2 spin liquid. Furthermore, we highlight the importance of assessing the subtle interplay of quantum and thermal fluctuations in reconciling the experimental findings on the nature of magnetic order in Yb$_2$Ti$_2$O$_7$.
\end{abstract}

\maketitle

\section{Introduction} 

In the broad landscape of quantum many-body systems, Mott-insulating magnets have long stood out as a versatile and materials-based platform for studying novel quantum phases. These range from conventionally ordered phases that nevertheless exhibit intricate spin textures such as helices~\cite{Binz-2006}, skyrmions~\cite{Muhlbauer-2009,Nagaosa-2013}, hedgehogs~\cite{Fujishiro-2019}, or platonic noncoplanar structures~\cite{Messio-2011,Gembe-2023,Gembe-2024}, to the scenario of a spin-nematic state~\cite{Andreev-1984} in which a magnet realizes the analog of a liquid crystal. In even more unconventional  cases, quantum spin-liquid phases~\cite{Savary-2017}  are stabilized, which cannot be characterized by symmetry-breaking order parameters and feature emergent degrees of freedom. 
Indeed, one of the recurring and probably most fascinating themes in condensed matter physics is the characterization of many-body phenomena involving an effective description of the low-energy behavior through emergent degrees of freedom. In some of its most interesting realizations, these emergent quasi-particles carry quantum numbers that are a fraction of those carried by the original degrees of freedom. A classic example of such a scenario is the appearance of phonons and rotons in a superfluid, where the emergent degrees of freedom are constrained by a set of conservation laws, which are different from and independent of the form of interaction between the atoms. Such an effective description has gained special attention in the context of magnetic systems and the stabilization of spin-liquid phases whose ground-state manifolds are identified as the set of ground states whose emergent degrees of freedom fulfill certain conservation laws. For magnetic lattices composed of a set of corner-sharing motifs, this description has been crucial in identifying and characterizing the distinct types of classical and quantum spin liquids that prevail in these geometries~\cite{Chalker2011}.

In this context, the pyrochlore lattice comprised of corner-sharing tetrahedra is an excellent platform for the study of a variety of classical and quantum spin liquids, both theoretically and experimentally. Among oxide compounds with a pyrochlore magnetic sublattice~\cite{Rau2019ARCMP,Hallas-AnnRevCMP}, conventional magnetic order~\cite{ross2011,Savary_QoBD_ErTiO,Yahne_enentrance_ErSnO,Sarkis_YbGeO,Hallas-AnnRevCMP},
classical spin liquids~\cite{Castelnovo2008}, 
and, possibly, quantum spin liquids~\cite{Smith_CeZrO} 
are realized. 
One paradigmatic classical spin liquid phase is ``spin ice", the ground state of the antiferromagnetic nearest-neighbor Ising model~\cite{Harris-1997,Castelnovo2008,Isakov-2004,Henley-2005}, in which the low-energy degrees of freedom are subject to an energetic constraint, 
the ``two-in-two-out" rule -- whose analogy to water ice has led to the term ``spin ice". 
This local constraint can be expressed as a 
Gauss' law on an emergent vector {\it gauge} field, i.e. $\partial_\alpha B^\alpha=0$ with $\alpha\in\{x,y,z\}$, defined on the links of the parent diamond lattice~\cite{Isakov-2004,Henley-2005}. This mathematical construction identifies spin ice as a rank-1 U(1) spin liquid (named after the rank-1 vector gauge field $B^\alpha$ fulfilling Gauss' law and consequently having an emergent U(1) symmetry) whose low-energy emergent gauge fields exhibit dipolar correlations  \cite{Henley-2010}. The dipolar correlations between these emergent fields result in the observation of anisotropic features in the spin correlation functions in reciprocal space known as twofold pinch points~\cite{Henley-2010,Castelnovo2008,Isakov-2005}. Fluctuations away from this ground-state manifold are understood as local violations of the Gauss' law constraint $\partial_\alpha B^\alpha=\rho$ and correspond to nonvanishing gauge charges in the system~\cite{Castelnovo2008}. 

Recent works~\cite{lozano_2024_atlas,lozano_2024arxiv,chung2023Arxiv2formu1spinliquids} have further extended the diversity of classical spin liquids observed for this lattice geometry. In particular, for the most general nearest-neighbor bilinear exchange Hamiltonian, the authors of Ref.~\cite{lozano_2024_atlas} identified and classified {\it all} possible classical spin liquids realized on the pyrochlore lattice. Such an analysis demonstrated that the emergent degrees of freedom describing the ground-state manifold are not only rank-1 fields~\cite{Castelnovo2008,lozano_2024_atlas,KTC_2024_phase,Henley-2010} of the form $B^\alpha$ (as introduced above), but can also be higher-rank fields~\cite{Francini2024nematicR2,benton2016} of the form $B^{\alpha\beta}$, or even a combination of the two~\cite{lozano-2023}. All of these classical spin liquids are U(1) spin liquids that have a set of associated Gauss' laws characterizing their ground-state manifold~\cite{lozano_2024_atlas} and are identified as \emph{algebraic} classical spin liquids within recently developed classification schemes~\cite{davier2023, yan2024,Fang-2023}.
Within this family of algebraic classical spin liquids, special attention has been devoted~\cite{Pretko-2017,Prem-2018,Yan-2020} to higher-rank spin liquids whose Gauss' law takes the form
\[
	\partial_{\alpha} B^{\alpha\beta}=0 \quad {\rm or} \quad \partial_\alpha \partial_\beta B^{\alpha\beta}=0 \,. 
\]
The higher-rank gauge theories associated with these Gauss's laws underlie the low-energy physics of so-called fracton spin liquids~\cite{Nandkishore-2019} which feature excitations with restricted mobility \textendash\ a consequence of the conservation of multipole moments of the gauge charges~\cite{Pretko-2017}. 

Although it has been noted, of late, that there exists a plethora of {\it classical} spin liquids realized in the pyrochlore lattice with a rich variety of emergent tensor gauge theories~\cite{lozano_2024_atlas,KTC_2024_phase}, much less is known about the quantum counterparts of these spin liquids,
partly because frustrated three-dimensional quantum magnets remain largely inaccessible to state-of-the-art numerical quantum many-body approaches. This paucity has, to some degree, been filled by the pseudo-fermion and pseudo-Majorana functional renormalization group (pf-FRG and pm-FRG, respectively) approaches~\cite{Muller-2024} which have enabled various forays into the three-dimensional world, unveiling the magnetic correlation profiles of models~\cite{lozano-2023,niggemann2023,Noculak-2023,Hering-2022,Kiese-2022,Iqbal-2019} and materials~\cite{gonzalez-2023,Ghosh-2019b,Chillal-2020,Iqbal-2017,Smith_CeZrO}. 
With the landscape of classical pyrochlore spin liquids mapped out~\cite{lozano_2024_atlas,KTC_2024_phase}, one might thus turn to these FRG approaches to explore the impact of quantum fluctuations on these classical spin liquids as parent states of novel quantum phases. One  pressing issue is the fate of the classical algebraic spin liquids once quantum fluctuations are introduced. Indeed, away from the classical limit of $S\to\infty$, quantum fluctuations may lead to tunneling between the degenerate states that span the ground-state manifold of a classical spin liquid~\cite{Hermele-2004}. These fluctuations 
are indispensable to realize a quantum spin liquid descending from a parent classical spin liquid \textendash\ the U(1) quantum spin ice serving as the quintessential case in point~\cite{Benton_seeing}. Furthermore, the introduction of quantum fluctuations not only modifies the effective theories describing a spin liquid~\cite{Benton_seeing} but may also reshape its immediate vicinity~\cite{lozano-2023}, possibly leading to the stabilization of novel exotic phases that have not been observed in the classical models.

In this manuscript, we study the phase diagram of the most general $S=1/2$ bilinear nearest-neighbor exchange Hamiltonian on the pyrochlore lattice by applying a fully generalized pf-FRG approach. We focus, in particular,  on the vicinity of a classical higher-rank spin liquid, the so-called pinch-line spin liquid~\cite{benton2016}. This choice is further motivated by the variety of synthesized pyrochlore compounds that have been found in the vicinity of this point~\cite{scheie2020,Yahne_enentrance_ErSnO,Savary_QoBD_ErTiO}. 
We demonstrate that the introduction of quantum fluctuations results in a substantial overall shift of the classical phase boundaries and the appearance of an extended nonconventional phase in parameter space where no conventional (dipolar) magnetic order is detected. Surprisingly, this nonconventional phase is {\sl not} centered around the classical triple point of maximum phase competition, see Figs.~\ref{fig:phasediagram-pinch-line}(a) and (b), thus defying conventional expectations, and highlighting the complete failure of linear spin-wave theory~\cite{yan2017}. 
Within the nonconventional phase, we identify different regimes which -- based on the analysis of spin structure factors -- we suggest host a quantum analog of the pinch-line spin liquid as well as a spin-nematic phase over a certain range of exchange parameters. The latter is remarkable as it provides a rare scenario of a quantum spin-nematic state (i) in the absence of a magnetic field and (ii) in three spatial dimensions, hitherto unreported for spin-$1/2$. 
The positioning of many rare-earth pyrochlore oxides in our quantum phase diagram brings the world of spin nematics within realistic material realizations. Our work also presents a unique example where a quantum order-by-disorder selection (beyond linear spin-wave treatment~\cite{yan2017}) at the classical triple point collapses the system into a unique ground state, thereby inducing conventional magnetic ordering, while thermal order-by-disorder fails to do so forming a classical spin liquid~\cite{benton2016}.

The rest of the paper is organized as follows: in Sec.~\ref{sec:ham}, we introduce the Hamiltonian and the irreducible representations of the single-tetrahedron point group $T_d$ used to classify the ordered phases and construct the emergent gauge fields used to construct the low-energy theory describing the spin liquid at the classical triple point. The model's classical phase diagram is summarized in Sec.~\ref{sec:classpd}. In Sec.~\ref{sec:pinchline}, we present the spin-$1/2$ quantum phase diagram obtained using pf-FRG, compare and contrast the quantum and classical phase diagrams, and identify a region in the quantum phase diagram where the quantum analog of the pinch-line spin liquid can be realized. In Sec.~\ref{sec:noncon}, we discuss in further detail the extended nonconventional region and further characterize some of the possible nonconventional phases within, such as the putative quantum spin nematic phase. In Sec.~\ref{sec:yb2}, we discuss the implications of our quantum phase diagram for $\rm Yb_2 Ti_2 O_7$, a compound that previous studies have found to be located near a phase boundary between two conventional magnetic orders~\cite{ross2011,robert2015,thompson2017,scheie2020}. Lastly, in Sec.~\ref{sec:outlook}, we summarize our findings and discuss further avenues for exploration to address the open questions. 

\section{Model}

\subsection{Hamiltonian and irreducible representations}
\label{sec:ham}

\begin{figure*}
    \centering
    \includegraphics{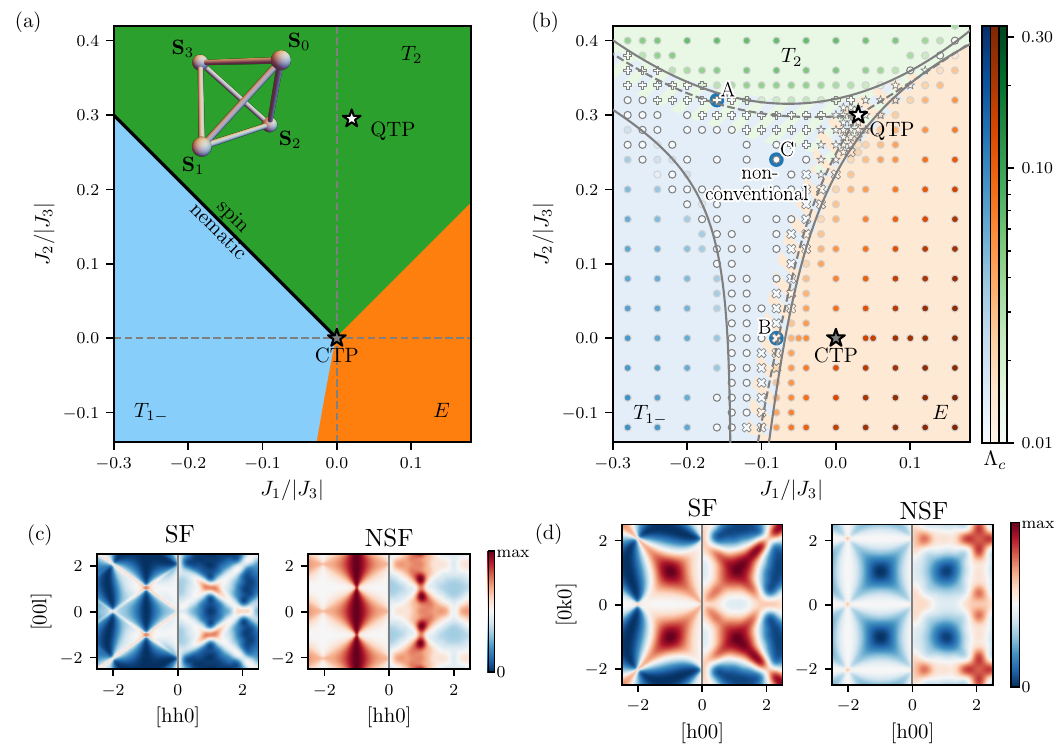}
    \caption{{\bf Classical and quantum phase diagram and neutron-scattering structure factors of the pinch-line spin liquid.} (a) Exact classical phase diagram (for $J_3 < 0, J_4 = 0$) showing three $\bm{q} = 0$ phases meeting at the classical triple point (CTP) at $J_1 = J_2 = 0$, where the ground state forms a classical pinch-line spin liquid. (b) Corresponding quantum phase diagram from pf-FRG. Background colors indicate the dominant order-parameter susceptibility, where hatched regions mark points where multiple order-parameter susceptibilities are nonvanishing. Different markers represent the number and type of nonvanishing susceptibilities: `$\circ$' for a single susceptibility, `+' for $T_{1-} \oplus T_2$, `x' for $E\oplus T_{1-}$ and `$\largestar$' for $E\oplus T_{1-} \oplus T_2$. The dashed lines are guides to the eye, highlighting where the dominant order-parameter susceptibility changes. These meet at the ``quantum triple point'' (QTP) at $J_1 \approx 0.03$, $J_2 \approx 0.3$, where all three susceptibilities are maximally degenerate. In conventionally ordered phases, the color saturation inside the ``o" markers (quantified by the three colorbars) reflects the critical scale $\Lambda_c$ at which a flow breakdown occurs, signaling the onset of dipolar magnetic order. In the ``nonconventional" phase (outlined by gray lines as the approximate phase boundaries), no flow breakdown occurs, indicating the absence of conventional magnetic order. Structure factors in the nonconventional phase at the QTP and the points A, B and C are shown in Fig.~\ref{fig:disordered_flows_and_sf}.
    (c) Neutron-scattering structure factors (see Sec.~\ref{sec:method} for the definitions) in the spin-flip (SF) and non-spin-flip (NSF) channel in the classical model at the CTP (left) from SCGA and in the quantum model at the QTP (right) from pf-FRG in the $[hhl]$-plane. (d) The same structure factors in the $[hk0]$-plane.}
    \label{fig:phasediagram-pinch-line}
\end{figure*}

Our starting point is the most general symmetry-allowed nearest-neighbor Hamiltonian on the pyrochlore lattice, which can be written in the form
\begin{align}
    &&H=\sum_{\langle ij \rangle}
    \bigg[ 
    J_{zz} \tilde{S}^z_i \tilde{S}^z_j - J_{\pm} (\tilde{S}^+_i \tilde{S}^-_j + \tilde{S}^-_i \tilde{S}^+_j)  \nonumber \\
    && + J_{\pm \pm} ( \gamma_{ij} \tilde{S}^+_i \tilde{S}^+_j + \gamma^{\ast}_{ij} \tilde{S}^-_i \tilde{S}^-_j)
    \nonumber \\
    && - J_{z \pm} ( \gamma^{\ast}_{ij} \tilde{S}^z_i \tilde{S}^+_j + \gamma_{ij} \tilde{S}^z_i \tilde{S}^-_j +  i \leftrightarrow j)
    \bigg] \,,
    \label{eq:hamiltonian}
\end{align}
where the sum is over nearest neighbors, $\tilde{S}_i^z$ is the local-$z$ spin degree of freedom oriented along the local $C_3$ axis, $\tilde{S}_i^\pm$ are the raising and lowering operators, and $\gamma_{ij}$ are bond-dependent phase factors~\cite{Rau2019ARCMP} (refer to Appendix~\ref{sec:irrep_decomposition} for complete definitions). In this Hamiltonian, the classical spin-ice spin liquid is realized for $J_{zz}>0$ while all other couplings are set to zero. Furthermore, recent works~\cite{lozano-2023,XXZ_PMFRG_nematic_2024arxiv} demonstrated that for non-Kramers pyrochlores for which $J_{z\pm}=0$, the introduction of quantum fluctuations by small nonvanishing $J_{\pm}$ and $J_{\pm\pm}$ interactions does not destroy the spin liquid phase. Indeed, in the limit $|J_{\pm}|,|J_{\pm\pm}|\ll J_{zz}$ the quantum spin-ice spin liquid is realized, while for $J_{\pm}\sim1/6$ and $J_{\pm\pm}\sim \pm 1/3$ a mixed rank-1 and rank-2 U(1) spin liquid is observed~\cite{lozano-2023}. In this work, we consider pyrochlore materials where the ions are in a Kramers-doublet crystalline-electric-field ground state, leading to a generally nonzero  $J_{z\pm}$, which prevents the emergence of the conventional spin-ice phase~\cite{Savary-2012}.

As a consequence of the corner-sharing structure of the pyrochlore lattice, the Hamiltonian in Eq.~\eqref{eq:hamiltonian} can be rewritten as a sum over tetrahedra $t$ as
\begin{equation}
    H = \sum_{\langle ij \rangle}  
    \bm S_i J_{ij} \bm S_j = \sum_{t} H^\mathrm{tet}[t] \,,
    \label{eq:matrix_hamiltonian}
\end{equation}
with the Hamiltonian on a single tetrahedron
\begin{equation}
    H^\mathrm{tet} = \frac{1}{2}\sum_{\mu,\nu\in t} \mathbf{S}_\mu \cdot \mathbf{J}_{\mu\nu} \cdot \mathbf{S}_\nu \,,
\end{equation}
where $\mu$ and $\nu$ label the sublattice structure of a single tetrahedron, $\mathbf{J}_{\mu\nu}$ is the corresponding spin exchange interaction between the two sublattices and $\mathbf{S}_\mu = (S^x_\mu, S^y_\mu, S^z_\mu)$ are spin-operators in the global frame, related to $\tilde{\mathbf{S}}_\mu$ by a basis transformation (see Appendix~\ref{sec:irrep_decomposition} for details). In particular, for the zeroth and first sublattice, see Fig.~\ref{fig:phasediagram-pinch-line}(a), the spin exchange matrix takes the form
\begin{equation}
    \mathbf{J}_{01} = 
    \begin{pmatrix}
        J_2 &J_4 &J_4 \\
        -J_4 &J_1 &J_3 \\
        -J_4 &J_3 &J_1
    \end{pmatrix},
    \label{eq:coupling-matrix-01}
\end{equation}
where $\{J_1,J_2,J_3,J_4\}$ are spin exchange couplings in the \emph{global} Cartesian basis, associated with the local spin exchange couplings $\{J_{zz},J_\pm,J_{\pm\pm},J_{z\pm}\}$ by a local rotation~\cite{yan2017}. All other exchange matrices can be obtained via the application of point group symmetry operations \cite{yan2017}. For conventional magnetic order, the classical ground state of this Hamiltonian is described by a $\bm{q}=0$ order~\cite{yan2017}, i.e., each tetrahedron basis displays the same spin order on each of the sublattices. It is, therefore, possible to restrict the classification of the ordered states in how these break the point group symmetry of a single tetrahedron, $T_d$.
To this end, we introduce the order parameters $\mathbf{m}_\lambda$ associated with the irreducible representations (irreps) $\lambda = \left\{A_2, E, T_{1-},T_{1+}, T_2\right\}$ of $T_d$. The order parameters $\mathbf{m}_\lambda$ are linear combinations of the Cartesian spin components $S_\mu^\alpha$ with $\alpha\in\{x,y,z\}$ within a single tetrahedron \cite{yan2017}, allowing the single tetrahedron Hamiltonian to be rewritten as
\begin{equation}
\begin{aligned}
     H^\mathrm{tet} = \frac{1}{2}\left[
    a_{A_2} m_{A_2}^2 
    + a_E\mathbf{m}_E^2 
    + a_{T_2}  \mathbf{m}_{T_2}^2  \right.\\ \left.
    + a_{T_{1-}} \mathbf{m}_{T_{1-}}^2 
    + a_{T_{1+}} \mathbf{m}_{T_{1+}}^2\right] \,.
\end{aligned}
\label{eq:irrep-decomposition}
\end{equation}
For details on the irrep decomposition and  definitions we refer the reader to Appendix \ref{sec:irrep_decomposition} as well as Refs.~\cite{yan2017,lozano_2024_atlas,KTC_2024_phase}. 
The classical ground state is determined by calculating the irrep with the minimal 
prefactor $a_\lambda$. In the case in which multiple $a_\lambda$ parameters are minimal, nonconventional magnetic phases such as spin nematics or nonmagnetic states such as spin liquids may be obtained. 

\subsection{Classical phase diagram}
\label{sec:classpd}

The resulting classical phase diagram for fixed $J_3 < 0$ and $J_4 = 0$ as obtained in Ref.~\cite{yan2017} is shown in Fig.~\ref{fig:phasediagram-pinch-line}(a) where three conventional magnetically ordered phases are indicated, namely an $E$, a $T_{1-}$ and a $ T_{2}$ phase. Right at the boundary between the $T_{1-}$ and the $ T_2$ phase it has been shown, based on classical Monte Carlo simulations~\cite{Francini2024nematicR2}, that the system exhibits spin nematicity, signaled by the onset of a quadrupolar order parameter. The remaining two phase boundaries in this phase diagram feature a thermal order-by-disorder~\cite{Noculak-2023} selection to a conventional $\bm q=0$ state. Lastly, the three magnetically ordered phases meet at an isolated point $J_1 = J_2 = J_4 = 0$ and $J_3<0$. We refer to this as the ``classical triple point'' (CTP), which in Fig.~\ref{fig:phasediagram-pinch-line} is marked by a gray star. At the CTP, extensive classical Monte-Carlo calculations suggest that no particular state is selected out of the degenerate ground-state manifold via a possible thermal order-by-disorder mechanism, therefore yielding a magnetically disordered state down to the $T\to 0$ limit. This implies the realization of a classical spin liquid (CSL), the so-called pinch-line spin liquid~\cite{benton2016}. The classical ground-state manifold for this CSL is defined by the constraints
\begin{equation}
    m_{A_2} = 0, \quad \mathbf{m}_{T_{1+}} = 0\label{eq:constraints}
\end{equation}
for every tetrahedron. These constraints do not fully determine an ordered ground state but leave the remaining fields $\mathbf{m}_{T_{1-}}, \mathbf{m}_{T_2}, \mathbf{m}_E$ to freely fluctuate. A soft spin treatment performed via a self-consistent Gaussian approximation (SCGA)~\cite{SCGA_Canals_pyrochlore} of the classical Hamiltonian reveals a flat band at the bottom of the spectrum, a consequence of the extensive ground-state degeneracy. More importantly, the band structure is gapless in a peculiar way: the first dispersive band touches the flat band not only at singular points but along a one-dimensional line in momentum space -- a nodal line. In more mathematical terms, this degenerate manifold can be described by a rank-2 field $B^{\alpha\beta}$ constructed from the $E$, $T_{1-}$ and $T_2$ irrep fields,
\begin{equation}
\begin{split}
      B^{\alpha\beta} =& \begin{pmatrix}
	2 m_E^1 & \sqrt{3} m_{T_2}^z    &    - \sqrt{3} m_{T_2}^y   \\
	- \sqrt{3} m_{T_2}^z  & -m_E^1 + \sqrt{3} m_E^2 &  \sqrt{3} m_{T_2}^x \\
	 \sqrt{3} m_{T_2}^y  &-\sqrt{3} m_{T_2}^x& -m_E^1 - \sqrt{3} m_E^2
\end{pmatrix}\\& - 
3\sin\theta
\begin{pmatrix}
	0 & m_{T_{1-}}^z &     m_{T_{1-}}^y \\
	 m_{T_{1-}}^z &0&   m_{T_{1-}}^x \\
	 m_{T_{1-}}^y &   m_{T_{1-}}^x & 0
\end{pmatrix},
\end{split}
\label{eq:rank2_PLSL}
\end{equation}    
where $\theta$ is a function of the coupling interaction parameters $\{J_1,J_2,J_3,J_4\}$, and $m_\lambda^\alpha$ are the components of the irrep fields; for more details we refer the reader to Appendix~\ref{sec:irrep_decomposition} and Ref.~\cite{lozano_2024_atlas}. In terms of this $B^{\alpha\beta}$ field, the constraints of Eq.~\eqref{eq:constraints} can be expressed as two Gauss's laws~\cite{benton2016,lozano_2024_atlas}, namely 
\[
	|\epsilon_{\alpha\beta\gamma}|\partial_\alpha B^{\beta\gamma}=0 \quad {\rm and} \quad \partial_\alpha B^{\alpha\beta}=0 \,,
\] 
where $\epsilon_{\alpha\beta\gamma}$ is the fully antisymmetric tensor. According to the classification of CSLs in Ref.~\cite{yan2024}, the ground state is therefore an algebraic nodal-line spin liquid. The gapless nature itself implies that all spin-spin correlations decay algebraically. Furthermore, the nodal line in the band structure results in a pinch-line singularity, a line in reciprocal space along which pinch-point features are observed~\cite{benton2016}. Indeed, within a soft-spin approximation, the resulting polarized neutron structure factor of this CSL, shown on the left-hand side of Fig.~\ref{fig:phasediagram-pinch-line}(c) and (d), exhibits twofold pinch points and pinch-lines along the $[111]$ and symmetry-related directions. We therefore refer to the CSL at hand as the {\sl classical pinch-line spin liquid} in the following.

\section{FRG results for the spin-$\mathbf{1/2}$ model}

\subsection{Fate of pinch-line spin liquid under quantum fluctuations}
\label{sec:pinchline}

We now turn to the question of how quantum fluctuations affect the classical pinch-line spin liquid. Indeed, the impact of quantum fluctuations on unconventional Coulomb spin liquids described by higher-rank gauge theories has recently been the subject of much debate. It has been argued in Ref.~\cite{lozano-2023} that some of the general features, such as conventional twofold pinch-points on the pyrochlore lattice observed in the classical model remain largely unchanged (with only quantitative modifications). In contrast, on the octochlore lattice, the pinch-lines (which are conventional
twofold pinch points in all planar cuts) are qualitatively affected while multifold pinch-points are completely washed out~\cite{niggemann2023}. 
Here, we study the effect of quantum fluctuations beyond linear spin-wave theory~\cite{yan2017} and show that a quantum analog of the pinch-line spin liquid can be stabilized for the spin-$1/2$ model; see Fig.~\ref{fig:phasediagram-pinch-line}(b).

To study the influence of quantum fluctuations on the classically observed phase diagram, we employ the pseudo-fermion functional renormalization group (pf-FRG) approach~\cite{Muller-2024}. It allows us to study the quantum model~\eqref{eq:hamiltonian} for spin $S = 1/2$ at zero temperature by the introduction of an infrared cutoff $\Lambda$, or \emph{RG scale}, into the theory. At high enough $\Lambda$, this corresponds to a high-temperature limit where all spins decouple and the correlations functions are known exactly. At $\Lambda \to 0$ the cutoff vanishes and the physical correlation functions are recovered. The interpolation between these regimes is governed by the FRG flow equations, an infinite hierarchy of differential equations for all correlation functions. We approximately solve these equations numerically, setting correlations beyond a bond distance $L$ to zero -- typically we consider up to $L = 7$ corresponding to 864 lattice sites, resulting in a total of $2.7 \times 10^7$ coupled differential equations that are integrated using HPC resources.
The main output of the pf-FRG is the flow of static spin-spin correlations of the form $\chi^{\alpha\beta, \Lambda}_{ij}\sim\langle S^\alpha_i S^\beta_j\rangle|^\Lambda_{\omega=0}$. 
A \emph{divergence} (or ``kink") in the flow of the spin-spin correlations in momentum space at a finite critical scale $\Lambda_c$  signals the formation of conventional, dipolar magnetic order characterized by an order parameter that is linear in the spin operators~\cite{Muller-2024}. Conversely, the \emph{absence} of such a {flow breakdown} implies either a quantum disordered phase or a nonconventional magnetic phase. In the latter case, although spin rotation symmetry is spontaneously broken, the phase is described by an order parameter that is nonlinear in the spin operators such as those arising in spin-nematic phases. While such nematic orders are not directly captured by our truncation scheme of the flow equations~\cite{Reuther-2010}, their presence can sometimes be assessed within pf-FRG via a linear-response framework~\cite{Iqbal-2016}. 

In the case of a flow breakdown, we can determine the emergent order by calculating the flow of the order-parameter susceptibilities. For a cartesian component $\alpha$ of the order parameter associated with the irrep $\lambda$, the susceptibility is defined as
\begin{equation}
\begin{aligned}
    &\langle m^\alpha_\lambda(\bm{q}) m^\alpha_\lambda(-\bm{q})\rangle = \\
    &\frac{1}{N_{u.c}} \sum_{t, t'} \exp\left[-i \bm{q} 
    \left(\mathbf{r}_\mathrm{t} - \mathbf{r}_\mathrm{t'}\right)\right] \langle m^\alpha_\lambda(\mathbf{r}_\mathrm{t}) m^\alpha_\lambda(\mathbf{r}_\mathrm{t'}) \rangle \,,
    \end{aligned}
    \label{eq:order-parameter-susceptibilities}
\end{equation}
where the sum runs over the tetrahedron unit cells of the pyrochlore lattice and $\mathbf{r}_t$ denotes the position of the tetrahedron centers. The largest order-parameter susceptibility at the flow breakdown evaluated at $\bm{q} = 0$
hints at the low-temperature order; see Sec.~\ref{sec:method} for more details. In the case of nonconventional phases where a clear flow breakdown is absent, we can utilize the neutron-scattering structure factor channels in the limit $\Lambda \to 0^+$ to compare to results of the corresponding classical model and experiments. 

\begin{figure}[ht!]
    \centering
    \includegraphics{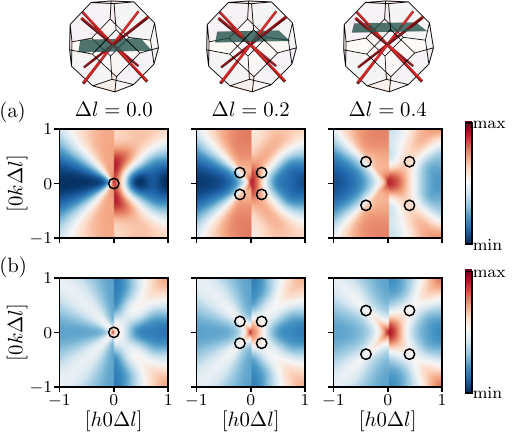}
    \caption{{\bf Indication of a pinch-line singularity in the order-parameter correlations}. (a) and (b) show the order parameter susceptibility $\langle m^1_E(\bm{q})m^1_E(-\bm{q})\rangle$ and $\langle m^z_{T_2}(\bm{q})m^z_{T_2}(-\bm{q})\rangle$, respectively.
    The left (right) side of each plot shows the correlation of the classical (quantum) model at the classical (quantum) triple point obtained via SCGA (pf-FRG). Each column shows a different horizontal plane inside the first Brillouin zone of the pyrochlore lattice, parameterized by $q_z = \Delta l$ and indicated in the schematic illustration above. As indicative of a pinch-line spin liquid, the correlations exhibit broadened pinch-point singularities where the planes of scattering cut the $[111]$ directions, highlighted by black circles in (a) and (b), and by red cylinders in the Brillouin zone illustration.}
    \label{fig:pinch-lines}
\end{figure}

\subsubsection*{Quantum phase diagram}

Employing pf-FRG we derived the quantum phase diagram in the vicinity of the CTP as shown in Fig.~\ref{fig:phasediagram-pinch-line}(b). Our pf-FRG calculations identify four distinct regimes; three conventionally ordered regimes, signaled by a flow breakdown in the spin-spin correlations and the strong rise of an irrep susceptibility, and a nonconventional regime without any apparent dipolar magnetic order (i.e.\ where the flow remains smooth down to the $\Lambda\to 0^+$ limit) in the center of the quantum phase diagram [indicated by the open symbols in Fig.~\ref{fig:phasediagram-pinch-line}(b), which denote the dominant short-range magnetic ordering pattern in the absence of true long-range order]. We provide a detailed description of the construction of the phase-diagram from the initial pf-FRG data in Appendix~\ref{app:pffrg-details}. 

Quantum fluctuations are expected to have their strongest impact in regions characterized by competing orders and extensive ground-state degeneracies. A natural starting point for analyzing their effects is therefore at the CTP. In contrast to the classical model, we observe a clear divergence in the flow of the $\mathbf{m}_E$ susceptibility at the CTP, implying that quantum fluctuations stabilize an ordered $E$ phase (see also Fig.~\ref{fig:irrep-flows} in Appendix~\ref{app:pffrg-details}). This constitutes a rare case where quantum fluctuations select an ordered state out of a degenerate ground-state manifold whereas thermal fluctuations do not. Furthermore, and similar to the results obtained in Ref.~\cite{lozano_2024arxiv}, the quantum phase boundaries of the ordered phases are significantly shifted with respect to those of the classical model; see Fig.~\ref{fig:phasediagram-pinch-line}(a) and (b) for a side-by-side comparison. A shift of similar magnitude of the phase boundary between the $T_{1-}$ and $E$ phase was already observed within nonlinear spin-wave theory~\cite{rau19} and exact diagonalization (at $T = 0$), as well as a numerical linked-cluster computation (NLC), and high-temperature expansion (HTE) (at finite $T > 0$)~\cite{Jaubert2015YbTiO}.
We note that a comparable shift of all phase boundaries can be obtained in the classical model by adding a nonzero, negative antisymmetric exchange of $J_4/|J_3| \approx -0.13$. Although initially zero in the limit $\Lambda \to \infty$, such interactions can, in principle, become nonzero in the effective low-energy two-point vertex functions, as they are generated during the FRG flow. This mechanism could shift the phase boundaries in our pf-FRG calculations, favoring the $E$ phase even when the microscopic value of $J_4$ is zero. 

\subsubsection*{Quantum triple point}
 
 The significant shift of the phase boundaries also suggests a substantial displacement of the triple point where the three orderings meet. Indeed we can identify a quantum analog of the CTP, which we refer to as the \emph{quantum triple point} (QTP), by determining the couplings (in the $J_3 < 0, J_4 = 0$ plane) where the difference between the order-parameter susceptibility of the $E, T_{1-}$ and $T_2$ phase is minimal in the $\Lambda \to 0^+$ limit. This point is significantly shifted towards the 
 upper-right in the quantum phase diagram (versus the location of the CTP in the classical phase diagram), with $J_1/|J_3| \approx 0.03$ and $J_2/|J_3|\approx 0.3$,
 which also places the QTP at the upper-right boundary of the nonconventional regime in the center of the quantum phase diagram (discussed in the next section). 
 At this QTP, the polarized neutron-scattering structure factors closely resemble those of the classical pinch-line spin liquid, 
 see Figs.~\ref{fig:phasediagram-pinch-line}(c) and (d), exhibiting broadened pinch lines as well as twofold and fourfold pinch points.   Moreover, the correlation functions of the $m_E^1$ and the $m_{T_2}^z$ irreps, shown in Fig.~\ref{fig:pinch-lines}, 
 also display twofold and fourfold pinch points along the $\langle111\rangle$ directions consistent with the findings in Ref.~\cite{benton2016}.

These similarities suggest that the magnetically disordered phase realized at and around the QTP is the quantum analog of the classical pinch-line spin liquid, which is possibly described by an emergent higher-rank gauge field $B^{\alpha\beta}$ constrained by an emergent Gauss's laws $\partial_\alpha B^{\alpha\beta}=0$ leading to the observation of the twofold and fourfold pinch points. It is worth noting that the pinch lines, twofold, and fourfold pinch points are not sharp, as is the case for the classical spin liquid but they exhibit a finite broadening. Such a broadening was also observed when studying other models~\cite{niggemann2023,Iqbal-2019} and was associated with quantum fluctuations causing local violations of the energetically imposed Gauss's law implying nonvanishing gauge charge fluctuations. 

\subsection{nonconventional phases}
\label{sec:noncon}

\begin{figure}[ht!]
    \centering
    \begin{overpic}{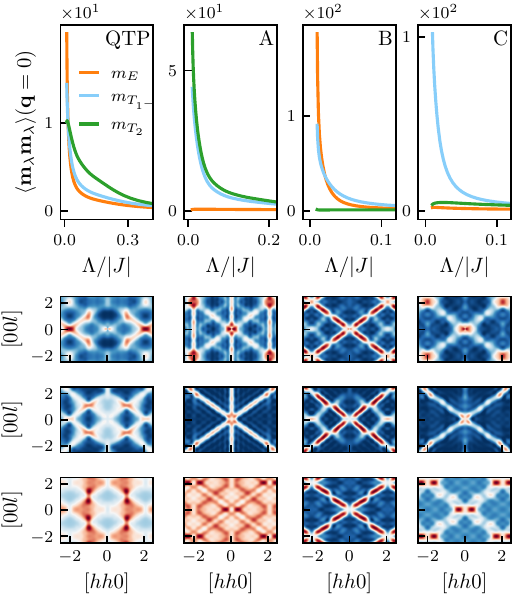}
    \end{overpic}
    \caption{{\bf Order-parameter susceptibility flows and structure factors in the nonconventional regime} for the couplings indicated in Fig.~\ref{fig:phasediagram-pinch-line} (QTP, A, B, C). The top panel shows the RG flow of the order-parameter susceptibility $\langle \mathbf{m}_\lambda \cdot \mathbf{m}_\lambda \rangle$ for the relevant irreps $\lambda \in \{E, T_{1-}, T_2\}$ (where $|J|^2=J_1^2 + J_2^2 + J_3^2 + J_4^2$ is used as a normalization). The smooth flow down to very low RG scales $\Lambda\to0$ indicate the lack of conventional magnetic order. The relative magnitude of the susceptibilities indicate which type of correlations dominate/compete.
    The bottom three rows show the 
    structure factor in the total, spin-flip, and non-spin-flip channels, respectively. While at the QTP the correlations resemble those of the classical pinch-line spin liquid, moving away from this point the correlations drastically change. The parameters  considered are $(J_1, J_2)$ = $(0.03, 0.3), (-0.28, 0.36), (-0.08, 0.0), (-0.08, 0.24)$ for points QTP, A, B, and C, respectively, with $J_3 = -1.0$ and $J_4 = 0$. 
    }
    \label{fig:disordered_flows_and_sf}
\end{figure}

In the previous section,  we have established that the QTP is the quantum analog of the classical pinch-line spin liquid and sits at the boundary of the nonconventional region in the quantum phase diagram. This identification is mainly based on the overall structure of the distinct spin correlation functions. Indeed, as we move away from the QTP towards the center of the nonconventional region, the unpolarized and polarized neutron structure factors drastically change.
This is illustrated in Fig.~\ref{fig:disordered_flows_and_sf} which shows the order-parameter susceptibility flow for a set of representative points, labeled $\rm QTP$, $\rm A,\ B, $ and $\rm C$, and marked in Fig.~\ref{fig:phasediagram-pinch-line}(b), and the corresponding unpolarized and polarized neutron structure factors obtained at the lowest simulated $\Lambda$. Interestingly, in the $\Lambda\to 0^+$ limit, the nonvanishing order-parameter susceptibility  for each of these points is distinct. This behavior is highlighted in Fig.~\ref{fig:phasediagram-pinch-line}(b) by regions with a hatched background, which indicate points where, in addition to the dominant order-parameter susceptibility, at least one other susceptibility has a relative magnitude exceeding 20\% . 

Taken together, our FRG results for the nonconventional regime in Fig.~\ref{fig:phasediagram-pinch-line}(b) are consistent with (at least) two different scenarios: (i) The regime presents a single quantum spin-liquid phase, with short-range correlations quantitatively changing upon parameter variations. (ii) The regime contains one or more nonconventional symmetry-breaking orders with higher-order multipole order parameters, such as a spin nematic.
The present pf-FRG framework which only keeps track of two-spin correlation functions is not suited to distinguish these scenarios \cite{Iqbal-2016,Noculak-2023}, asking for complementary numerical techniques to settle this question.
Nevertheless, the similarities between the structure factors obtained from FRG and the ones of the classical model strongly suggest that scenario (II) applies. In what follows we present a careful discussion for various parameter sets within the nonconventional regime, based on the structure factors and the prevailing irreps, which support this view.

\begin{figure}
    \centering
    \includegraphics[width=\textwidth]{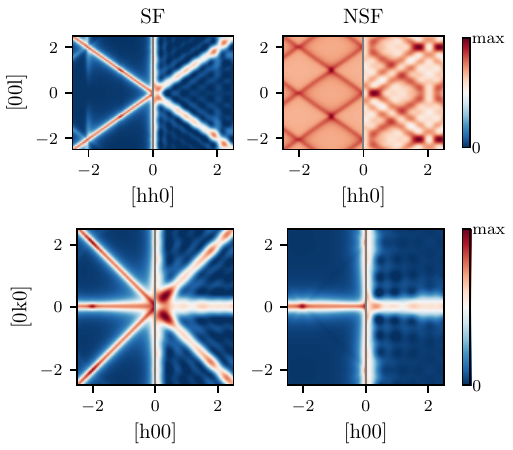}
    \caption{{\bf Structure factors of the putative spin nematic phase} in the $[hhl]$-plane (top) and the $[hk0]$-plane (bottom), as would be measured using polarized neutron scattering in the spin-flip (left) and non-spin-flip (right) channels. Each panel compares results from the classical model obtained via the SCGA (left) with those from the quantum model computed using the pf-FRG (right). The calculations are performed at $(J_1, J_2, J_3, J_4) = (-0.28,0.28,-1,0)$ (classical) and $(J_1, J_2, J_3, J_4) = (-0.28, 0.36, -1.0, 0.0)$ (quantum), corresponding to the phase boundary between the $T_{1-}$ and $T_2$ phases in the classical and quantum phase diagrams, as shown in Fig.~\ref{fig:phasediagram-pinch-line}.}
    \label{fig:nematic_phase}
\end{figure}

\subsubsection*{Spin nematics for the $T_{1-}\oplus T_2$ regime }
\label{sec:noncon-nematic}

The RG flow at the representative point $\rm A$ shows a quantitative degeneracy between the $T_{1-}$ and $ T_2$ fields. At this point, the unpolarized neutron structure factor exhibits continuous lines of scattering, similar to the pinch lines of the classical pinch-line spin liquid~\cite{benton2016}. In the classical model, the boundary between these two phases yields an unconventional magnetic order (invariant under time reversal) -- a spin nematic phase characterized by a quadrupolar order parameter~\cite{Francini2024nematicR2,Taillefumier_2017}. The degeneracy between the $T_{1-}$ and $T_2$ fields results in an accidental U(1) symmetry spanned by a set of single-tetrahedron spin configurations parameterized as $m^\alpha(\theta)=m_{T_2}^\alpha\cos(\theta)+m_{T_{1-}}^\alpha\sin(\theta)$~\cite{Noculak-2023,Francini2024nematicR2}. Furthermore, the band structure of the classical Hamiltonian Eq.~\eqref{eq:hamiltonian} along the $ T_{1-}\oplus  T_2$ line displays flat lines in the low-energy bands~\cite{KTC_2024_phase,Francini2024nematicR2}. The authors of Ref.~\cite{Francini2024nematicR2} showed that the correlation functions in this phase, henceforth referred to as the $T_{1-}\oplus T_2$ phase, display continuous lines of scattering similar to those observed in our pf-FRG calculations. The presence of these ``rods" is ascribed to the low-energy bands of the Hamiltonian which show flat lines along the $[111]$, $[001]$ and symmetry-related directions in reciprocal space. Along these lines, there occur band touchings at the $[hkl]=[000]$, $[hkl]=[111]$, and $[hkl]=[200]$ points; see Fig.~\ref{fig:bands_cuts} in Appendix \ref{sec:evolution_energy_bands}. These features in the energy contours result in the observation of high-intensity lines in the correlation functions in reciprocal space, produced by the flat lines, together with slightly higher intensity at the band touching points, at the classical level. 
Indeed, an SCGA analysis at intermediate temperatures of a point along the $ T_{1-}\oplus  T_2$ boundary, with $(J_1,J_2,J_3,J_4)=(-0.28,0.28,-1,0)$, yields similar structure factors with continuous lines of scattering as those observed at point $\rm A$; see Fig.~\ref{fig:nematic_phase}. At low temperatures the only intensity comes from the continuous line of scattering and no other features can be observed. We refer the reader to Appendix~\ref{section:SCGA_method} for a more detailed discussion of the temperature regimes studied in the SCGA. Our pf-FRG calculations find these continuous lines of scattering to be {\it robust} features in a narrow region along the $T_{1-}$\textendash$ T_2$ boundary in the quantum phase diagram [see the hatched region marked in Fig.~\ref{fig:phasediagram-pinch-line}(b)]. Consequently, we identify the point $\rm A$ and all those points where only the flow of the $T_{1-}$ and $T_2$ irreps is nonvanishing, i.e., belonging to the $T_{1-}\oplus T_2$ phase, with possible spin-nematic order in the quantum model. 
In particular, this implies that, unlike the classical model where the spin-nematic state is strictly located at a phase boundary, quantum fluctuations stabilize it over what appears to be a finite region in the quantum model, possibly indicating the existence of a nematic phase. We point out, however, that the rods primarily indicate the near-degeneracy of the two $T_{1-}$ and $ T_2$ irreps.

 Lastly, we note that the authors in Ref.~\cite{Francini2024nematicR2} showed that for the classical ($S\to \infty$) scenario this system exhibits a first-order transition driven by an order-by-disorder selection and characterized by a spin-nematic order parameter which is bilinear in the spin degrees of freedom. In this low-temperature phase, the spin configurations simultaneously break the accidental U(1) symmetry by selecting a set of angles $\{\theta\}$ in the U(1) manifold, as well as the $C_3$ cubic symmetry, where only certain Cartesian components of the $T_{1-}$ and $T_2$ irreps remain thermally populated~\cite{Francini2024nematicR2}. Hints of this cubic symmetry breaking are observed in the low-temperature structure factors where the intensity of the rods is no longer equivalent along different directions related by cubic symmetry.

\begin{figure}
    \centering
    \includegraphics{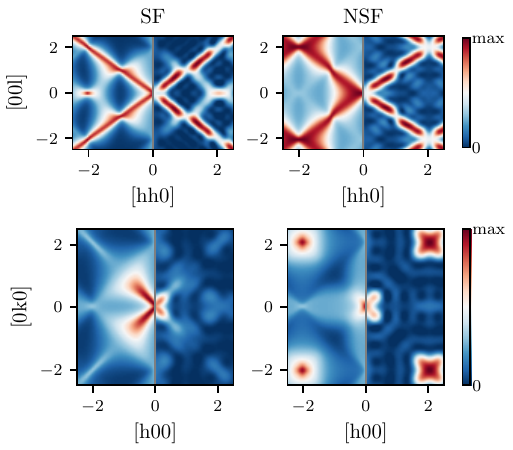}
    \caption{{\bf Structure factors of the spin model at the boundary between the $E$ and $T_{1-}$ phases} in the $[hhl]$-plane (top) and the $[hk0]$-plane (bottom), as would be measured using polarized neutron scattering in the spin-flip (left) and non-spin-flip (right) channels. Each panel compares results from the classical model obtained via the SCGA (left) with those from the quantum model computed using the pf-FRG (right). The calculations are performed at $(J_1,J_2,J_3,J_4) \approx (-0.0189, -0.1, -1.0,0.0)$ (classical) and $(J_1,J_2,J_3,J_4) = (-0.08, 0.0, -1.0, 0.0)$ [quantum, point $\rm B$ in Fig.~\ref{fig:phasediagram-pinch-line}(b)].}
    \label{fig:boundary_phase}
\end{figure}

\subsubsection*{Nonconventional $E\oplus T_{1-}$ regime}
\label{sec:noncon-e-plus-t1}

Moving on, for the representative point $\rm B$, a nonvanishing RG flow for the $E$ and $T_{1-}$ susceptibilities is found. In the classical model, a degeneracy between these two irreps occurs at their phase boundary. The pf-FRG correlation functions at this point show two important features: rods of scattering along the $[111]$ directions, and a lobe of intensity at the $[hkl]=[220]$ and symmetry-related points. These same features are observed in the structure factors of classical spin models located at the phase boundary and also between the $E$ and the $T_{1-}$ phases at intermediate temperatures where a cooperative paramagnetic regime is realized; see Fig.~\ref{fig:boundary_phase}. The origin of the rods of scattering lies in the low-energy bands of Hamiltonian being flat lines {\it only} along the $[111]$ and symmetry-related directions in reciprocal space. Additionally, along these lines, there are band touchings at the $[hkl]=[000]$ and $[hkl]=[111]$ points, hence, these are reflected in the structure factor (at the classical level) as flat lines with slight enhancement of intensity at the band-touching points. It is worth noting that the lobe of intensity observed at the $[hkl]=[220]$ point is not associated with band touchings with the flat lines but instead correspond to a minimum in the band structure where four bands meet; see Appendix~\ref{sec:evolution_energy_bands}. In the classical model, however, a symmetry-breaking transition takes place, selecting an $E$ phase at intermediate temperatures followed by a transition into the $T_{1-}$ phase at low temperatures~\cite{Jaubert2015YbTiO,Scheie_PRL_Dynamic_YbTiO,scheie2020}. Just as for the representative point $\rm A$, we observe absence of dipolar ordering tendencies in our pf-FRG calculations which could be indicative of a possible nonconventional order or a quantum spin liquid. We label this phase the $ E\oplus T_{1-}$ phase as, based on the RG flow of the order-parameter susceptibilities, configurations in this nonconventional phase would be composed by a mixture of $E$ and $T_{1-}$ fields. 

Indeed, such a proposal for low-temperature spin configurations was provided by the authors of Ref.~\cite{scheie2020} in the context of $\rm Yb_2 Ti_2 O_7$, a compound that lies in close proximity to the classical phase boundary between these two phases~\cite{ross2011,thompson2017,robert2015,scheie2020}. In that work, the authors concluded that the magnon spectrum for this compound is best represented by a mixture of these two irreps and was a clear signature of the strong competition between the two phases. We return to this point and discuss the experimental predictions of our results in the next section.

\subsubsection*{Nonconventional $T_{1-}$-only regime }
\label{sec:noncon-t1-only}

Lastly, we discuss the representative point $\rm C$. At this point, only the $T_{1-}$ order-parameter susceptibility grows to a significant magnitude. Nevertheless, our pf-FRG simulations detect no dipolar ordering tendencies. Up to the smallest infrared cutoff $\Lambda$, the structure factor for this point only shows diffuse features. These features are reminiscent of those observed in the representative points $\rm A$ and $\rm B$, and therefore of the $ E\oplus T_{1-}$ and spin-nematic $ T_{1-}\oplus T_2$ phase; see Fig.~\ref{fig:disordered_flows_and_sf}. Moreover, we note that the only common order-parameter susceptibility that plateaus to a nonvanishing value for these phases is precisely that of the $T_{1-}$ irrep. Altogether, these results suggest that the representative point $\rm C$ is located within an intermediate mixed phase between both the spin nematic $T_{1-}\oplus T_{2}$ and $E\oplus T_{1-}$ phases. Our overall analysis thus points to the fact that the nonconventional region in the spin-$1/2$ quantum phase diagram is composed of an ensemble of phases.

\section{Consequences for modeling pyrochlore materials}
\label{sec:yb2}

Our results for the quantum phase diagram identify an overall shift of the $\bm q=0$ phase boundaries with respect to those of the classical model as well as the emergence of nonconventional magnetic phases. These results are most relevant for those materials whose interaction parameters are located close to a classical phase boundary: materials that were associated with some magnetically ordered phase in the classical model may be associated with another phase in the quantum model. 

\subsubsection*{$\rm Yb_2Ti_2O_7$}

One example of such a material is $\rm Yb_2Ti_2O_7$~\cite{ross2011,thompson2017,robert2015,scheie2020}. Within a classical model description, there is a growing consensus that places this compound in a ferromagnetic $T_{1-}$ phase, however, in close proximity to an $E$ phase. This proximity leads to a plethora of interesting phenomena associated with the strong competition between these two phases~\cite{Jaubert2015YbTiO,robert2015,Scheie_PRL_Dynamic_YbTiO}. Indeed, a recent theoretical and experimental work found that the optimal low-temperature spin configurations of $\rm Yb_2Ti_2O_7$ used to reproduce the experimentally measured magnon spectra are mixed configurations of the $E$ and the $T_{1-}$ states~\cite{scheie2020}. However, this observation is not compatible with the classical ground state prediction where only a $T_{1-}$ phase should be observed at low temperatures. Indeed, in the classical model any $E$ spin configurations in a $T_{1-}$ phase correspond to excitations above the ground state. The disagreement between theoretical predictions and experimental observations may imply either that there exists an intermediate mixed phase not captured in the classical analysis, or that the determined interaction parameters of this compound have to be revisited.

To obtain the interaction parameters, the authors of Ref.~\cite{scheie2020} fit their high-field neutron-scattering data to spin-wave spectra. The resulting fit, however, is underconstrained: they identify a continous line in the parameter space spanned by $J_1, J_2, J_3$ and $J_4$ along which the spin-wave calculations reproduce the experimental spectra equally well. Similar observations were made in an earlier study~\cite{robert2015}. To constrain the fit, they additionally match the zero-field excitation gap predicted by their spin-wave calculations to the neutron-scattering data. This gap closes at the transition from the FM ($T_{1-}$) to the AFM ($T_2$) phase, making it a measure of proximity to the corresponding phase boundary. If, as our pf-FRG calculations suggest, quantum fluctuations significantly shift this phase boundary, then the zero-field gap will likely be shifted as well. This could, in turn, lead to a significant change in the best-fitting interaction parameters -- comparable in magnitude to the shift of the phase boundaries we observe in our calculations. Unfortunately, pf-FRG currently does not provide access to dynamic (real-frequency) correlation functions of spin-anisotropic models \cite{potten2025keldysh}, and therefore we cannot confirm this in a quantitative manner.

\begin{figure}[b]
    \centering
    \includegraphics{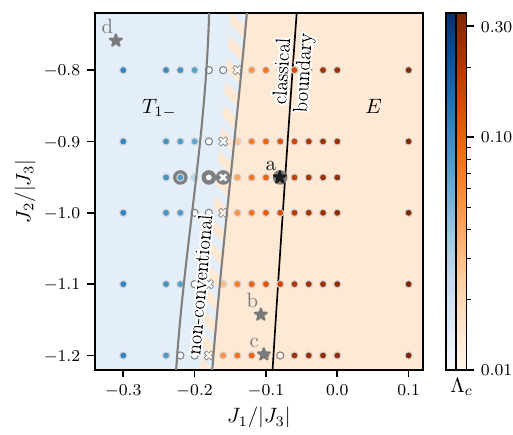}
    \caption{\textbf{Quantum phase diagram in the vicinity of Yb$_2$Ti$_2$O$_7$} from pf-FRG. The couplings $J_3$ and $J_4$ are fixed to the estimates for Yb$_2$Ti$_2$O$_7$ from Scheie \textit{et al.} \cite{scheie2020} (a). The stars show the estimated values of $J_1$ and $J_2$ for Yb$_2$Ti$_2$O$_7$ from (a) Scheie \cite{scheie2020}, (b)  Robert \cite{robert2015}, (c) Thompson~\cite{thompson2017} and (d) Ross~\cite{ross2011}. Note that the values of $J_3$ and $J_4$ for (b)–(d) differ from those in (a), so the corresponding parameters do not lie exactly in the plane shown in the figure (see Appendix~\ref{app:nlswt-comparison} for numerical values). Order-parameter susceptibility flows and structure factors for the four points underlaid by large gray circles are shown in Fig.~\ref{fig:structure-factor-ybti}.
    }
    \label{fig:phase-diagram-ybti}
\end{figure}

\begin{figure}[ht!]
    \centering
    \includegraphics{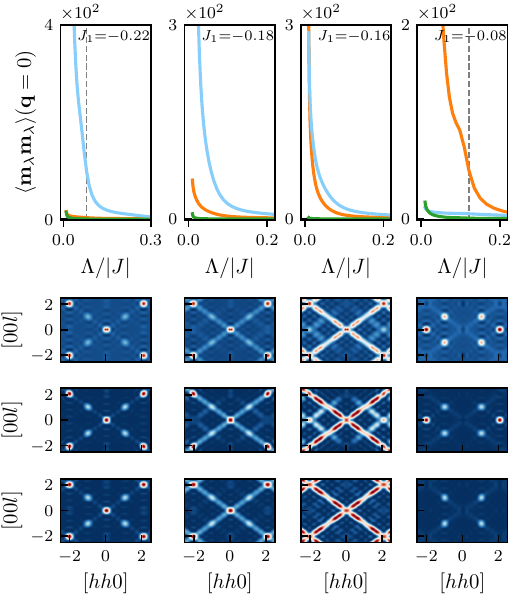}
    \caption{{\textbf{RG flows and structure factors along the transition from the $T_{1-}$ to $E$ phase} for the points underlaid by gray circles in Fig.~\ref{fig:phase-diagram-ybti}. For these points, $J_2, J_3$, and $J_4$ are fixed to the  Yb$_2$Ti$_2$O$_7$ parameters from Ref.~\cite{scheie2020}. The right-most panel ($J_1/|J_3| \approx -0.08$) corresponds to exact Yb$_2$Ti$_2$O$_7$ parameters. The top panel shows the RG flow of the order-parameter susceptibilities for the relevant irreps $\lambda \in \{E, T_{1-}, T_2\}$ (where $|J|^2=J_1^2 + J_2^2 + J_3^2 + J_4^2$ is used as a normalization). The dashed gray lines indicate the critical scale $\Lambda_c$ indicating the onset of conventional magnetic order. The bottom three panels show the total spin structure factor as well as the structure factors in the SF and NSF channel, respectively.}}
    \label{fig:structure-factor-ybti}
\end{figure}

We now revisit the quantum pyrochlore phase diagram in the vicinity of the boundary between the $E$ and the $T_{1-}$ phases, compare our results with other existing works studying the $S=1/2$ case, and assess the repercussion of these findings from a material perspective. Fig.~\ref{fig:phase-diagram-ybti} illustrates the quantum phase diagram obtained via pf-FRG for a plane in the interaction parameter space where the interaction parameters for $\rm Yb_2Ti_2O_7$ obtained by Ref.~\cite{scheie2020} lie. This quantum phase diagram displays an overall shift of the phase boundaries along with the appearance of a nonconventional magnetic regime. The evolution of the structure factors and the order-parameter susceptibilities along the phase transition from the $T_{1-}$ to the nonconventional and ultimately to the $E$ phase are shown in Fig.~\ref{fig:structure-factor-ybti} (for the exemplary points underlaid by gray circles in Fig.~\ref{fig:phase-diagram-ybti}). 

For large negative $J_1$, the susceptibility flow of $\mathbf{m}_{T_{1-}}$ is clearly dominant and exhibits a flow breakdown. Additionally, the structure factors feature sharp peaks at the $[000]$ and $[111]$ points, characteristic of a conventionally ordered $T_{1-}$ ferromagnetic phase. 
As $J_1$ increases, the flow breakdown disappears (see Appendix~\ref{app:pffrg-details} for details on our flow-breakdown criterion), signaling the emergence of a nonconventional phase. On the left side of this phase, the $T_{1-}$ susceptibility remains dominant, placing it in the nonconventional $T_{1-}$-only regime discussed in Sec.~\ref{sec:noncon-t1-only}. The structure factors already faintly display extended rods of scattering along the $[111]$-directions (and symmetry-related axes) in addition to the high-intensity lobes associated with the locations in reciprocal space where Bragg peaks related to a $T_{1-}$ symmetry-breaking phase would be stabilized. These rods fully manifest towards the right of the nonconventional phase, accompanied by additional peaks at $[220]$ and symmetry-related momenta. In this region (marked by ``x" markers and a hatched background color in Fig.~\ref{fig:phase-diagram-ybti}), both the $E$ and $T_{1-}$ susceptibility are of similar magnitude, placing it in the $E \oplus  T_{1-}$ regime discussed in Sec.~\ref{sec:noncon-e-plus-t1}. The structure factors in this regime are consistent with those measured in low-energy neutron-scattering experiments on $\rm Yb_2Ti_2O_7$ at finite but low temperatures above the critical temperature, where the material realizes a short-range correlated phase \cite{Scheie_PRL_Dynamic_YbTiO}. However, we observe the same behavior at zero temperature, suggesting that quantum fluctuations alone seem to allow the simultaneous stabilization of $T_{1-}$ and $E$ correlations. Upon further increasing $J_1$, the rods in the structure factor disappear, leaving only sharp peaks at $[220]$, $[111]$ and symmetry-related momenta~\cite{Sarkis_YbGeO}, while the $E$ susceptibility becomes dominant and exhibits a flow breakdown -- clear signatures of a conventionally ordered $E$ phase. 

Compared to the classical model, the $E$ phase is significantly enlarged, constituting a shift of the phase boundary by approximately $\Delta J_1 \approx -0.1/|J_3|$.
A shift of similar magnitude was already observed in the spin-$1/2$ model within exact diagonalization (at $T = 0$), as well as a numerical linked-cluster computation (NLC) and high-temperature expansion (HTE) (at high temperatures) \cite{Jaubert2015YbTiO}. Additionally, we note that the regime exhibiting $E$ order in the quantum but $T_{1-}$ in the classical model shows quantitative agreement with a region identified via nonlinear spin-wave theory in a previous study~\cite{rau19} where an instability of the $T_{1-}$ phase was observed; see Appendix~\ref{app:nlswt-comparison} for more details. A similar quantitative agreement between nonlinear spin-wave theory and pf-FRG was also reported in a recent study where an instability close to a quantum spin liquid phase was observed by both approaches~\cite{Hickey_2024arxiv}. This agreement further solidifies the plausibility of our findings and therefore supports the possible experimental observation of the nonconventional phase between the conventionally ordered $E$ and $T_{1-}$ phases. 
Interestingly, in the quantum phase diagram in Fig.~\ref{fig:phase-diagram-ybti}, the interaction parameters for $\rm Yb_2Ti_2O_7$ obtained by Ref.~\cite{scheie2020} locate this compound in the $E$ phase, as opposed to the $T_{1-}$ phase in the classical phase diagram. Although this would be inconsistent with the experimental observation which identifies a $T_{1-}$ phase at $T_c$, we reiterate that the phase diagram we have obtained is, strictly speaking, a $T=0$ phase diagram. It is therefore possible that thermal fluctuations lead to an invasion of the $T_{1-}$ over the $E$ phase at nonzero temperatures.
Indeed, a similar behavior is observed in the RG flow of the order-parameter susceptibilities for the estimated $\rm Yb_2Ti_2O_7$ parameters. For larger cutoffs $\Lambda/|J| > 0.3$ (within  a mean-field treatment the cutoff $\Lambda$ can be interpreted as temperature~\cite{Iqbal-2016}), the $T_{1-}$ susceptibility is the most prominent, while the $E$ susceptibility becomes clearly dominant only at lower cutoffs.  We refer the reader to Appendix~\ref{app:pffrg-details} for a more detailed discussion. Such a scenario is also observed for the classical model~\cite{Jaubert2015YbTiO,Scheie_PRL_Dynamic_YbTiO} where the long-range ordered phase at intermediate temperatures does not correspond to the $T\to0$ order. This discrepancy between the observed ordered phase at $T_c$ and the predicted ordered phase as $T\to0$ motivates further experimental investigation of this compound and theoretical studies that concomitantly account for quantum and thermal fluctuations at low temperatures. 

Altogether, our findings suggest that the ground state of $\rm Yb_2Ti_2O_7$ lies within the long-range ordered  $E$-phase (as opposed to what has been previously reported). 
We speculate that the proximity of this compounds to the $ E$-$ T_{1-}$ phase boundary may result in the observation of a finite-temperature phase 
where the $T_{1-}$ phase is dominant, but the $ E$ phase still prevails at sufficiently low temperatures -- a scenario proposed in a recent work investigating this compound~\cite{scheie2020}.

\section{Conclusion and Outlook}
\label{sec:outlook}
We have established the spin-$1/2$ quantum phase diagram along selected cuts of the most general symmetry-allowed nearest-neighbor Hamiltonian on the pyrochlore lattice, accurately treating quantum fluctuations within a pseudo-fermion functional renormalization group (pf-FRG) approach. The corresponding classical model hosts a triple point (between $E$, $T_{1-}$ and $T_2$ irrep magnetic orders) where thermal order-by-disorder fails to select a unique ground state and the system realizes a classical spin liquid characterized by pinch-line singularities in the spin structure factor reflective of a generalized rank-2 U(1) electromagnetism~\cite{Yan-2020}. In contrast to thermal effects, we find that quantum order-by-disorder selects the $E$-irrep  ordered state at the parametric location of the classical triple point, thus presenting a rare example where quantum fluctuations stabilize an ordered phase and thermal fluctuations do not. Nevertheless, we do find an appreciable region in parameter space where conventional (dipolar) magnetic order is absent but, contrary to conventional expectations, this region is not centered around the classically degenerate triple point or phase boundaries. Our results thus call into question the validity of a linear spin-wave treatment of quantum fluctuations for this model which reported the absence of magnetic order at the triple point and in a region centered around it. 

In the nonconventional magnetic region of our quantum phase diagram, we identify a parameter set where the susceptibility of the $E$, $T_{1-}$ and $T_2$ irreps becomes degenerate and consequently, there emerges a quantum triple point. Notably, this degeneracy is found to persist over an extended region in parameter space. Based on a careful assessment of spin structure factors, we find evidence for the appearance of a higher-rank gauge theory in the nonmagnetic phase not only at the quantum triple point but over an extended region. Deciphering the precise microscopic nature of this unconventional phase at and around the quantum triple point, which could putatively be a quantum spin liquid with nontrivial gauge groups, would constitute an important future endeavor. This should involve a projective symmetry-group classification of gapless and gapped mean-field Ans\"atze with different low-energy gauge groups and subsequently assessing the impact of gauge fluctuations through Gutzwiller projection within a variational Monte Carlo approach. 

Another intriguing aspect of the classical phase diagram is the presence of a spin nematic phase along the phase boundary of the $T_{1-}$ and $T_2$ magnetic phases. Its presence is revealed in the spin structure factor by the presence of rod-like features along given directions together with lobes of intensity at particular positions. Interestingly, we find these fingerprints of the spin-nematic phase in the nonconventional region of our spin-$1/2$ quantum phase diagram, in a sliver of parameter space along the boundary with the $T_2$ phase. The spin-nematic is an elusive state and its existence in three spatial dimensions and in the absence of a magnetic field has been contested~\cite{Sato-2013}. Thus, it would be important to further substantiate our findings employing complementary many-body approaches which can directly access the quadrupole order parameter correlation functions to establish their long-range behavior. Furthermore, the regions near the boundaries of the nonconventional phase with the $E$ and $T_{1-}$ phases present qualitatively distinct structure factors, thus pointing to yet another distinct nonconventional phase, whose nature remains {\it terra incognita}. Our work thus highlights the rich composition of the nonconventional region of the quantum phase diagram as being built out of a variety of novel quantum phases, and sets the stage for future works aimed at identifying their precise nature.

We show that the opening of an appreciable window of a nonconventional phase together with a significant parametric shift of the phase boundaries for the spin-$1/2$ quantum has important implications for understanding the physics of pyrochlore oxides, in particular {\rm Yb$_2$Ti$_2$O$_7$}. The fact that  experimental observation point to the onset of a $T_{1-}$ order at $T_{c}$, while the previously estimated model Hamiltonian parameters based on the classical model~\cite{ross2011,thompson2017,robert2015,scheie2020} place the material within the E phase of our $T=0$ quantum phase diagram, highlights a need to better understand the intertwined effect of quantum and thermal fluctuations for this material. We argue here that thermal fluctuations could cause subtle shifts in the phase boundary at $T\neq0$ between $E$ and $T_{1-}$ so as to locate Yb$_2$Ti$_2$O$_7$ in the $T_{1-}$ or $ E\oplus T_{1-}$ phases, which would then be consistent with experimental observations. The joint impact of quantum and thermal fluctuations could potentially be investigated within the recently developed pseudo-Majorana functional renormalization group approach~\cite{Niggemann-2021}, and would constitute an important future line of investigation towards reconciling theory and experiment on Yb$_2$Ti$_2$O$_7$. In similar vein, it is important to note that sample quality for {\rm Yb$_2$Ti$_2$O$_7$}~\cite{Rau2019ARCMP,Hallas-AnnRevCMP} has evolved over the years and this would presumably affect the thermodynamic behavior of a given sample specimen under investigation.

 We note that, in principle, the pf-FRG can also be formulated for arbitrary spin-lengths $S>1/2$ \cite{baez2017}. However, it has so far been implemented and tested only for spin models with diagonal Heisenberg interactions of the form $J_{ij} \mathbf{S}_i\mathbf{S}_j$. This approach  involves an artificial enlargement of the Hilbert space, assuming the ground state remains unchanged and the unphysical states do not affect the FRG flow. Although adding level repulsion terms to the Hamiltonian that favor the physical subspace support this assumption for Heisenberg models, its validity for anisotropic pyrochlore Hamiltonians studied here but with $S>1/2$ remains unclear. Although beyond the scope of this work, confirming this method for non-Heisenberg models would enable the systematic study of the transition from the classical $S\to\infty$ to the quantum $S\to 1/2$ limit. This would prove to be important in studying materials with spin moments $S>1/2$ such as NaCaNi$_2$F$_7$~\cite{Plumb2019_NaCaNiO}, which we leave as an investigation for future work.\\[8mm]


\section{Methods}
\label{sec:method}

\subsection{Pseudo-fermion functional renormalization group}
As outlined in the main text, applying the pf-FRG requires numerically solving the  flow equations for the correlation functions of interest. To this end, we have extended the pf-FRGSolver.jl Julia package \cite{pffrgsolver} to accommodate models featuring arbitrary nondiagonal components in the spin interactions matrices. For an efficient implementation and feasible run times, this extension involved a proper utilization of combined real-space and spin-space symmetries present in the model at hand~Eq.~\eqref{eq:hamiltonian}. The package already provides state-of-the-art integration routines for the pf-FRG flow equations within the Katanin truncation \cite{katanin2004} at zero temperature, $T = 0$. The ordinary differential equations are solved using the Bogacki–Shampine method -- a third-order Runge-Kutta method with adaptive step-size control. To capture the dependence on three continuous Matsubara frequencies of the four-point correlation functions (the fourth frequency being fixed by energy conservation), the package utilizes discrete adaptive frequency grids and multilinear interpolation to obtain off-grid values. In our simulations, we use a frequency grid comprising $N_\omega = 35$ discrete bosonic frequencies and $N_\nu \times N_\nu = 30 \times 30$ fermionic frequencies. To simulate an infinite lattice, correlations beyond a certain  bond-distance $L$ are set to zero. Unless otherwise specified, we typically use system sizes of $L = 3, 5, 7$, with larger sizes $L = 9$ employed for specific points of interest. A more detailed discussion of our implementation can be found in Ref.~\cite{Kiese-2022}. Details about the method itself, its capabilities, and caveats are described in a recent review article~\cite{Muller-2024}.

The primary output of our pf-FRG calculations is the flow of the static ($\omega = 0$) spin-spin correlations
\begin{equation}
    \label{eq:pf-FRG-correlations}
    \chi_{ij}^{\Lambda \alpha\beta} = \int_0^\infty\!\!\! d\tau e^{i\omega \tau}\! \left\langle \hat{T}_\tau S^\alpha_i(\tau)S^\beta_j(0)\right\rangle\Big|^\Lambda_{\omega = 0} \ \,
\end{equation}
where $\hat{T}_{\tau}$ denotes the time-ordering operator in imaginary time $\tau$, and $i,j$ are arbitrary sites on the pyrochlore lattice. A straight forward Fourier transform then yields the corresponding spin-spin correlations in momentum space. If the ground state of a model spontaneously breaks symmetry captured by an order-parameter linear in the spin-operators, the relevant components of the spin-spin correlations will, in theory, diverge at a finite critical scale $\Lambda_c > 0$ at certain momenta $\bm {q}_\mathrm{max}$ depending on the type of order. In practice, due to the approximations applied to the flow equations, this divergence may soften to a cusp or a kink, which becomes more pronounced as the lattice size $L$ increases. 

In the absence of a distinct divergence, there is no definitive criterion for unequivocally identifying such flow breakdowns from numerical data. Additionally, the pf-FRG has shown a tendency to overestimate the extent of disordered regions in parameter space. To address this, we perform a detailed analysis of both the flow of the correlations and their second derivatives, identifying nonmonotonic behavior in the second derivative that intensifies with increasing L as indicative of a flow breakdown. For examples, we refer the reader to Appendix~\ref{app:pffrg-details}. We note that the phase boundaries between conventionally ordered and ``nonconventional" regime in Fig.~\ref{fig:phasediagram-pinch-line}(b) should not be considered quantitatively precise. These boundaries depend on the criterion used for identifying flow breakdowns and the details of the numerical implementation. Nevertheless, the qualitative features -- such as the existence and approximate location of the nonconventional phase -- should be largely robust to these variations. Additionally, numerically resolving the flow of correlations becomes significantly more challenging at lower $\Lambda$. This may explain the seemingly more diffuse phase boundary between the $T_{1-}$ and the nonconventional phase in Fig.~\ref{fig:phasediagram-pinch-line}(b), near which critical scales below $\Lambda/|J| < 0.02$ appear. We have drawn the approximate phase boundaries as our best estimate beyond which our numerical data show no evidence of a flow breakdown.

To further classify the nature of a phase, we can directly calculate the order-parameter susceptibilities, defined in Eq.~\eqref{eq:order-parameter-susceptibilities} (with the $\mathbf{m}_\lambda$ given in Appendix~\ref{sec:irrep_decomposition}), as well as the neutron-scattering structure factors from the Fourier-transformed  spin-spin correlations
\begin{equation}
    \chi^{\Lambda\alpha\beta}_{\mu\nu}(\bm{q}) = \langle S_{\mu}^\alpha(\bm{q}) S_{\nu}^\beta(-\bm{q})\rangle^\Lambda = \sum_{i \in \mu}\sum_{j \in \nu} e^{-i\bm{q} (\mathbf{r}_i - \mathbf{r}_j)} \chi_{ij}^{\Lambda\alpha\beta},
\end{equation}
where $\mu,\nu \in (1,2,3,4)$ label the sublattices of the pyrochlore lattice, the sums $i \in \mu$ and $j \in \nu$ run over all sites in the respective sublattices, and $\mathbf{r}_i$ denote their positions. For unpolarized neutrons, the structure factor then takes the form
\begin{equation}
	{S}_\perp(\bm{q}) = \sum_{\alpha,\beta}
    \sum_{\mu,\nu}
    \left(\delta_{\alpha,\beta} -\hat{\bm{q}}^\alpha \hat{\bm{q}}^\beta\right)\chi^{\alpha\beta}_{\mu\nu}(\bm{q}) \,,
\end{equation}
where we don't denote the $\Lambda$-dependence for brevity. On the other hand, the polarized neutron structure factors are defined in terms of the incident neutrons' polarization $\hat{z}_{\rm N}$~\cite{Chung_flatband}, effectively separating the unpolarized neutron structure factor into two channels, namely
the non-spin-flip (NSF) channel 
\begin{equation}
	\label{eq:nsf-neutron}
	 {S}_\perp^{\mathrm{NSF}}(\bm{q}) = \sum_{\alpha,\beta}
     \sum_{\mu,\nu}
     \left(\hat{z}_{\textrm N}^\alpha \hat{z}_{\textrm N}^\beta\right) \chi^{\alpha\beta}_{\mu\nu}(\bm{q})
\end{equation}
and the spin-flip channel  
\begin{equation}
	\label{eq:sf-neutron}
 	{S}_\perp^{\mathrm{SF}}(\bm{q}) = {S}_a\perp (\bm{q})-{S}_\perp^{\mathrm{NSF}}(\bm{q}) \, .
\end{equation}
In the above equations, we have assumed for simplicity that the magnetic moments $\bm \mu_i$ directly correspond to the spin moments, i.e. $\bm \mu_i^\alpha = g_i^{\alpha\beta}\bm S_i^\beta=\bm S_i^\alpha$ where the $g$-tensor is taken to be isotropic. All momentum resolved structure factors shown in this manuscript are calculated at a minimal cutoff of $\Lambda/|J| = 0.02$ in the nonconventional phase (i.e. in the absence of a flow breakdown), or close to the critical scale $\Lambda_c$ (right at the flow breakdown) in the conventionally ordered phases.

\subsection{Self-consistent Gaussian approximation}

The self-consistent Gaussian approximation (SCGA), often referred to as the large-${N}$ approximation~\cite{SCGA_Canals_kagome,SCGA_Canals_pyrochlore,lozano-2023}, is a classical approximation where the hard spin-length constraint, $|\bm S_i|^2\equiv S^2$, is replaced by the soft-spin constraint, $\frac{1}{N}\sum_i|\bm S_i|^2\equiv S^2$ with $N$ being the number of spins in the system, where this constraint is satisfied on average. To enforce this constraint, we introduce a Lagrange multiplier $\lambda$ which is obtained self-consistently for each temperature
\begin{equation}
\frac{1}{N}\sum_{m,\bm{q} }\left(\frac{\varepsilon_m(\bm{q})}{T}+\lambda\right)^{-1}\equiv S^2, \label{eq:lagrange_multiplier_constraint}
\end{equation}
where the $\varepsilon_m(\bm{q})$ are the eigenenergies of the $12\times12$ matrix $J_{ij}$ in Eq.~\eqref{eq:matrix_hamiltonian}. The resulting theory yields a Gaussian theory that can be solved exactly and from which all spin correlation functions can be computed from the general correlation function 
\begin{equation}
    \chi^{\alpha\beta}_{\mu\nu}(\bm{q})=\langle S_{\mu}^\alpha(\bm{q}) S_{\nu}^\beta(-\bm{q})\rangle =\left( \frac{\bm{J}_{\mu\nu}^{\alpha\beta}(\bm{q})}{T}+\lambda \right)^{-1}\, ,\label{eq:SCGA-chi}
\end{equation}
where the indices $\mu,\nu$ label the sublattice index, while the sub-indices $\alpha,\beta$ label the spin components. From this susceptibility, the same observables as in the pf-FRG approach can be straightforwardly computed, including the order-parameter correlations and the neutron-scattering structure factors defined above.


\vspace{4mm}
\noindent{\bf \raggedright Data availability} \\
The numerical data shown in the figures and the raw FRG data is available on Zenodo~\cite{gresista_zenodo}.
\\

\noindent{\bf \raggedright Acknowledgments} \\
%
We thank Karlo Penc, Bernhard Wortmann, Sid Para\-meswaran, Johannes Reuther, and Michel J. P. Gingras for discussions.
The Cologne group acknowledges partial funding from the DFG within Project ID No. 277146847, SFB 1238 (Projects C02 and C03). L.G. thanks IIT Madras for a IoE Visiting Graduate student fellowship during which this project was initiated.  
D.L.-G. and MV acknowledge financial support from the DFG
through the W\"urzburg-Dresden Cluster of Excellence on Complexity
and Topology in Quantum Matter -- \textit{ct.qmat} (EXC 2147, project-id
390858490) and through SFB 1143 (project-id 247310070). D.L.-G. is
supported by the Hallwachs-R\"ontgen Postdoc Program of \textit{ct.qmat}.
The work of S.T. and Y.I. was performed in part at the Aspen Center for Physics, which is supported by National Science Foundation Grant No.~PHY-2210452 and a grant from the Simons Foundation (1161654, Troyer). This research was supported in part by Grant No. NSF PHY-2309135 to the Kavli Institute for Theoretical Physics (KITP).
Y.I. acknowledges support from the ICTP through the Associates Programme, from the Simons Foundation through Grant No.~284558FY19, IIT Madras through the Institute of Eminence (IoE) program for establishing QuCenDiEM (Project No.~SP22231244CPETWOQCDHOC), and the International Centre for Theoretical Sciences (ICTS), Bengaluru  during a visit for participating in the program: Kagome off-scale (ICTS/KAGOFF2024/08).
The numerical simulations were performed on the JUWELS cluster at the Forschungszentrum J\"ulich 
and the Noctua2 cluster at the Paderborn Center for Parallel Computing (PC$^2$).
Y.~I.~acknowledges the use of the computing resources at HPCE,  IIT Madras.

\appendix

\section{Hamiltonian in the global and local frame}

In our definition of the Hamiltonian both in the local frame Eq.~\eqref{eq:hamiltonian} and the global frame Eq.~\eqref{eq:matrix_hamiltonian} we primarily follow the conventions of Ref.~\cite{yan2017}, but restate all relevant definitions here for completeness. 

The basis sites of the tetrahedral unit cell shown in Fig.~\ref{fig:phasediagram-pinch-line}(a), are defined relative to its center as
\begin{equation}
    \begin{aligned}
    \mathbf{r}_0 = \frac{a}{8} \left(1, 1, 1\right) \quad
    \mathbf{r}_1 = \frac{a}{8} \left(1,-1,-1\right) \\
    \mathbf{r}_2 = \frac{a}{8} \left(-1,1,-1\right) \quad
    \mathbf{r}_3 = \frac{a}{8} \left(-1,-1,1\right) 
    \end{aligned} \,,
\end{equation}
where $a$ is the lattice spacing. The Hamiltonian Eq.~\eqref{eq:hamiltonian} is defined in the local frame, where the local spin $\tilde{\mathbf{z}}$-axis of $\tilde{S}_i$ aligns with the vector connecting the tetrahedron center to the corresponding basis site $\mathbf{r}_i$. This direction corresponds to the local $\langle 111 \rangle$ axis with $C_3$ symmetry. The local $\tilde{\mathbf{x}}$ and $\tilde{\mathbf{y}}$ axes are chosen following the convention introduced in Ref.~\cite{ross2011} in which all local $\tilde{\mathbf{y}}$-axes lie in the same plane. In this convention, the spin operators in the global frame $\mathbf{S}_\mu$ and the local frame $\tilde{\mathbf{S}}_\mu$ are related by the basis transformation 
\begin{equation}
    \tilde{\mathbf{S}}_\mu = \mathbf{R}_\mu \mathbf{S}_\mu \,,
\end{equation} 
with the rotation matrices $\mathbf{R}_\mu$ for the basis sites $\mu = 0, 1, 2, 3$ defined as
\begin{equation}
\begin{aligned}
    \mathbf{R_0} = \frac{1}{\sqrt{6}} \begin{pmatrix}
       -2 & 1 & 1 \\
        0 & -\sqrt{3} & \sqrt{3} \\
        \sqrt{2} & \sqrt{2} & \sqrt{2}
    \end{pmatrix} \, ,\\
        \mathbf{R_1} = \frac{1}{\sqrt{6}} \begin{pmatrix}
        -2 & -1 & -1 \\
        0 & \sqrt{3} & -\sqrt{3} \\
        \sqrt{2} & -\sqrt{2} & -\sqrt{2}
    \end{pmatrix}  \, ,\\
    \mathbf{R_2} = \frac{1}{\sqrt{6}} \begin{pmatrix}
       2 & 1 & -1 \\
        0 & -\sqrt{3} & -\sqrt{3} \\
        -\sqrt{2} & \sqrt{2} & -\sqrt{2} \\
    \end{pmatrix}  \, , \\
    \mathbf{R_3} = \frac{1}{\sqrt{6}} \begin{pmatrix}
        2 & -1 & 1 \\
        0 & \sqrt{3} & \sqrt{3} \\
        -\sqrt{2} & -\sqrt{2} & \sqrt{2} 
    \end{pmatrix} \, .
\end{aligned}
\end{equation}
The rows of the matrices correspond to the local $\tilde{\mathbf{x}}, \tilde{\mathbf{y}}$ and $\tilde{\mathbf{z}}$ axes, respectively. The bond-dependent phase factors $\gamma_{ij}$ appearing in the Hamiltonian in the local frame are
\begin{equation}
    \gamma = \begin{pmatrix}
        0 & 1 & -e^{-i\pi/3} & -e^{i\pi/3} \\
        1 & 0 & -e^{i\pi/3} & -e^{-i\pi/3} \\
        -e^{-i\pi/3} & -e^{i\pi/3} & 0 & 1 \\
        -e^{i\pi/3} & -e^{-i\pi/3} & 1 & 0 \\
    \end{pmatrix} \, .
\end{equation}
The coupling matrices $\mathbf{J}_{\mu\nu}$ of the Hamiltonian in the global frame \eqref{eq:matrix_hamiltonian} can be obtained using the basis transformation stated above, and then collecting the terms coupling the spins on basis sites $\mu$ and $\nu$. This, for example, leads to the coupling matrix $\mathbf{J}_{01}$ defined in Eq.~\eqref{eq:coupling-matrix-01}. The coupling matrices are additionally related by the symmetries of the pyrochlore lattice. Concretely, this means they can, e.g., be constructed from $\mathbf{J}_{01}$ by a $C_3$ rotation around the local $\tilde{\mathbf{z}}$-axis of basis site $\mu = 0$ and/or a $C_2$ rotation around the global $\mathbf{z}$-axis through the center of the tetrahedra
\begin{equation}
    \mathbf{C_3} = \begin{pmatrix}
        0 & 0 & 1 \\
        1 & 0 & 0 \\
        0 & 1 & 0
    \end{pmatrix}\, ,\quad
    \mathbf{C_2} = 
    \begin{pmatrix}
        -1 & 0 & 0 \\
        0 & -1 & 0 \\
        0 & 0 & 1
    \end{pmatrix} \, .
\end{equation}\\
Combined with a possible lattice inversion along the corresponding bond (represented by a matrix transpose) the remaining coupling matrices are then related to $\mathbf{J}_{01}$ by the symmetry transformations that map the corresponding bonds onto each other, namely
\begin{equation}
    \begin{aligned}
    \mathbf{J}_{02} &= \mathbf{C_3} \mathbf{J}_{01} \mathbf{C_3}^T \, ,\\
    \mathbf{J}_{03} &= \mathbf{C_3}^T \mathbf{J}_{01} \mathbf{C_3} \, , \\
    \mathbf{J}_{12} &= \mathbf{C_3}^T \mathbf{C_2}^T \mathbf{J}_{01}^T \mathbf{C_2} \mathbf{C_3} \, ,\\
    \mathbf{J}_{13} &= \mathbf{C_3} \mathbf{C_2}^T \mathbf{J}_{01}^T \mathbf{C_2} \mathbf{C_3}^T \, ,\\
    \mathbf{J}_{23} &= \mathbf{C_2} \mathbf{J}_{01}^T \mathbf{C_3}^T \, ,
    \end{aligned}
\end{equation}
which leads exactly to the matrices stated in Ref.~\cite{yan2017}. 

The relation between the coupling parameters typically used in the local and global frame is given by
\begin{align}
\label{eq:global-to-local}
    \begin{pmatrix}
        J_{zz} \\ J_{\pm} \\ J_{\pm\pm}\\J_{z\pm}
    \end{pmatrix}
    = \frac{1}{6}
    \begin{pmatrix}
        -4 & 2 & -4 & -8 \\
         2 & -1 & -1 & -2 \\
         1 & 1 & -2 & 2 \\
         \sqrt{2} & \sqrt{2} & \sqrt{2} & -\sqrt{2} \\
    \end{pmatrix}
    \begin{pmatrix}
        J_1\\J_2\\J_3\\J_4
    \end{pmatrix},
\end{align}
allowing a straightforward comparison between the two frames. 

There is a duality in the global parametrization: A rotation by $\pi$ around the local $\tilde{\mathbf{z}}$-axes leads to $\tilde{S}^\pm \to - \tilde{S}^\pm$ and $\tilde{S}^z \to +\tilde{S}^z$, effectively mapping $J_{z\pm} \to -J_{z\pm}$~\cite{Rau2019ARCMP}. The corresponding dual global frame, which we parameterize by $(\bar{J}_1, \bar{J}_2, \bar{J}_3, \bar{J}_4)$ strongly mixes the exchange constants of the original global frame as
\begin{equation}
\label{eq:dual-parameters}
\begin{pmatrix}
    \bar{J}_1 \\ \bar{J}_2 \\ \bar{J}_3 \\ \bar{J}_4
\end{pmatrix}
=\frac{1}{9}
\begin{pmatrix}
    5 & -4 & -4 & 4 \\
-8 & 1 & -8 & 8 \\
-4 & -4 & 5 & 4 \\
2 & 2 & 2 & 7
\end{pmatrix}
\begin{pmatrix}
    J_1 \\ J_2 \\ J_3 \\J_4
\end{pmatrix}.
\end{equation}
Although not used in the main text, we will use this parametrization in order to compare to other literature results in Appendix~\ref{app:nlswt-comparison}.

\begin{figure}
    \centering
    \begin{overpic}[width=\textwidth]{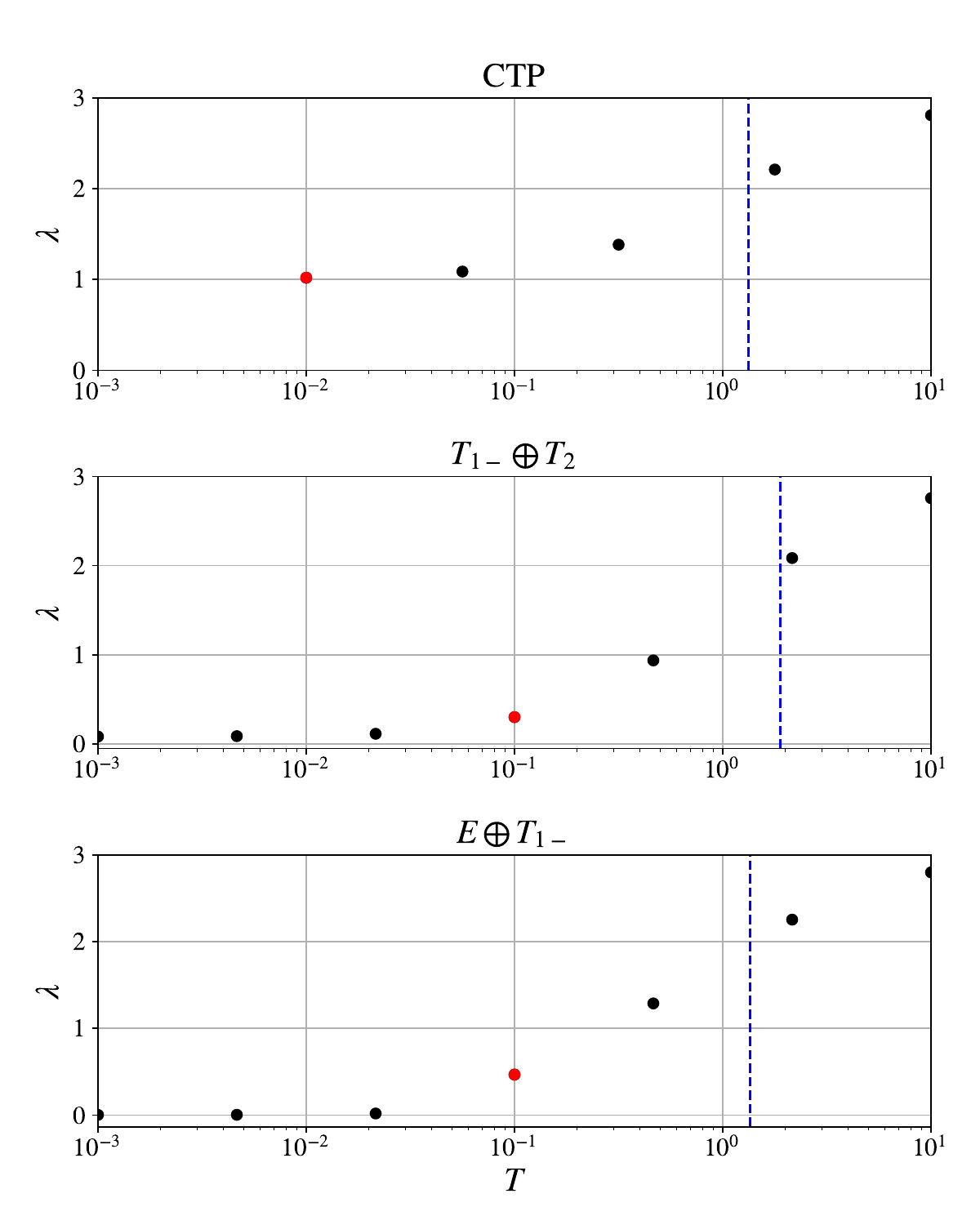}
    \put(10,88){(a)}
    \put(10,58){(b)}
    \put(10,26){(c)}
    \end{overpic}
    \caption{{\bf Evolution of the Lagrange multiplier $\lambda$ as a function of temperature.} Here panels (a), (b), and (c) displays the evolution of $\lambda$ for the set of interaction parameters used in Figs.~\ref{fig:phasediagram-pinch-line}, \ref{fig:nematic_phase}, and \ref{fig:boundary_phase}, respectively. The blue dashed line in each panel corresponds to $T_{\mathrm{PM}}^{\mathrm{MFT}}$. The red dot in these panels indicates the temperature at which the data shown in Figs.~\ref{fig:phasediagram-pinch-line}, \ref{fig:nematic_phase}, and \ref{fig:boundary_phase} are obtained.}
    \label{fig:lambda_T}
\end{figure}

\section{Irrep decomposition and order-parameter fields}
\label{sec:irrep_decomposition}

The order-parameter fields used in the irrep decomposition of the classical Hamiltonian in Eq.~\eqref{eq:irrep-decomposition} are defined in terms of spin operators in the global frame, following Ref.~\cite{yan2017}, as 
\begin{widetext}
\begin{equation}
  \begin{aligned}
    m_{A_2} &= \frac{1}{2 \sqrt{3}}\left(S_0^x+S_0^y+S_0^z+S_1^x-S_1^y-S_1^z-S_2^x+S_2^y-S_2^z-S_3^x-S_3^y+S_3^z\right) \, ,\\
    \mathbf{m}_E &= \begin{pmatrix}
        \frac{1}{2 \sqrt{6}} \left(-2 S_0^x+S_0^y+S_0^z-2 S_1^x-S_1^y-S_1^z+2 S_2^x+S_2^y-S_2^z+2 S_3^x-S_3^y+S_3^z\right)\\
        \frac{1}{2 \sqrt{2}}\left(-S_0^y+S_0^z+S_1^y-S_1^z-S_2^y-S_2^z+S_3^y+S_3^z\right)
    \end{pmatrix} \, ,\\
    \mathbf{m}_{T_{1A}} &= \begin{pmatrix}
        \frac{1}{2}\left(S_0^x+S_1^x+S_2^x+S_3^x\right) \\
        \frac{1}{2}\left(S_0^y+S_1^y+S_2^y+S_3^y\right) \\
        \frac{1}{2}\left(S_0^z+S_1^z+S_2^z+S_3^z\right)
    \end{pmatrix} \, , \\
    \mathbf{m}_{T_{1B}} &= \begin{pmatrix}
        \frac{-1}{2 \sqrt{2}}\left(S_0^y+S_0^z-S_1^y-S_1^z-S_2^y+S_2^z+S_3^y-S_3^z\right) \\
        \frac{-1}{2 \sqrt{2}}\left(S_0^x+S_0^z-S_1^x+S_1^z-S_2^x-S_2^z+S_3^x-S_3^z\right) \\
        \frac{-1}{2 \sqrt{2}}\left(S_0^x+S_0^y-S_1^x+S_1^y+S_2^x-S_2^y-S_3^x-S_3^y\right)
    \end{pmatrix}\, , \\
    \mathbf{m}_{T_{2}} &= \begin{pmatrix}
        \frac{1}{2 \sqrt{2}}\left(-S_0^y+S_0^z+S_1^y-S_1^z+S_2^y+S_2^z-S_3^y-S_3^z\right) \\
        \frac{1}{2 \sqrt{2}}\left(S_0^x-S_0^z-S_1^x-S_1^z-S_2^x+S_2^z+S_3^x+S_3^z\right) \\
        \frac{1}{2 \sqrt{2}}\left(-S_0^x+S_0^y+S_1^x+S_1^y-S_2^x-S_2^y+S_3^x-S_3^y\right)
    \end{pmatrix}\, , \\
    \mathbf{m}_{T1-} &= \cos{\theta} \mathbf{m}_{T_{1A}} - \sin{\theta} \mathbf{m}_{T_{1B}}\, ,\\
    \mathbf{m}_{T1+} &= \sin{\theta} \mathbf{m}_{T_{1A}} + \cos{\theta} \mathbf{m}_{T_{1B}}\, ,
  \end{aligned}
\end{equation}
\end{widetext}
where the angle
\begin{equation}
    \theta = \frac{1}{2} \arctan \left(\frac{\sqrt{8} J_3}{2 J_1+2 J_2+J_3-2 J_4}\right)
\end{equation}
is chosen such that the nonzero coupling between the original $\mathbf{m}_{T_{1A}}$ and $\mathbf{m}_{T_{1B}}$ fields is removed for $\mathbf{m}_{T_{1-}}$ and $\mathbf{m}_{T_{1+}}$. Physically, $\theta$ represents the canting angle between the spins in the $T_1$ ground state, which form a splayed ferromagnet around the $\langle111\rangle$ (or a symmetry-related) axis. The prefactors $a_\lambda$ of the order-parameter fields in the Hamiltonian are given by
\begin{eqnarray}
        a_{A_2} 	& = & -2J_1 + J_2 - 2J_3 + 4J_4 \,, \nonumber \\
        a_{E} 	& = & -2J_1 + J_2 + J_3 + 2J_4 \,, \nonumber \\
        a_{T_{1-}} & = & (2 J_1+J_2) \cos^2\theta -(J_2+J_3-2 J_4) \sin^2\theta \nonumber \\
        && +\sqrt{2} J_3 \sin2\theta \,, \nonumber \\
        a_{T_{1+}} & = & (2 J_1+J_2) \sin^2\theta  -(J_2+J_3-2 J_4) \cos^2\theta \nonumber \\
        && -\sqrt{2} J_3 \sin2\theta \,, \nonumber \\
        a_{T_2} 	& = & -J_2 + J_3 - 2J_4 \,.
\end{eqnarray}
Each order parameter can reach a maximum value of $\mathbf{m}_\lambda^2 =1$. The spin-length constraint for classical $S = 1/2$ spins $\mathbf{S}_i^2=1/4$ implies $\sum_\lambda \mathbf{m}_\lambda^2 = 1$. Therefore, the order-parameter field with the lowest prefactor $a_\lambda$ exactly determines the classical ground state, which is then of $\bm{q}=0$ type. More exotic ground states can only be realized at phase boundaries and critical points, where multiple $a_\lambda$ are degenerate, as is the case for the nonconventional phase discussed in the main text. 

\section{Application of the SCGA at low, intermediate, and high temperatures}
\label{section:SCGA_method}

In the main text, the SCGA is used to predict the evolution of the correlation functions for the spin systems that we have considered. Although previous works demonstrated that this approximation accurately describes the thermal evolution of spin correlations for a variety of pyrochlore systems~\cite{benton2016,Yan-2020,lozano-2023,lozano_2024_atlas}, as this theory is purely quadratic, it is incapable of properly capturing a phase transition to a symmetry-breaking phase in classical models. However, within this theory, the temperature evolution of the Lagrange multiplier $\lambda$ serves as a qualitative indicator of the realization of an ordered phase (at low temperatures), a cooperative-paramagnetic phase (at intermediate temperatures), and paramagnetic phases (at high temperatures). Assuming that the interaction matrix $ \bm{J}^{\alpha\beta}_{\mu\nu}(\bm{q})$ has no low-energy flat bands, the Lagrange multiplier $\lambda$ approaches $0$ in the low-temperature regime, $n=3$ (where $n$ is the number of spin components) in the high-temperature regime, and an intermediate value, i.e. $0<\lambda<3$, in the remaining temperature regime.  In the case in which the interaction matrix $ \bm{J}^{\alpha\beta}_{\mu\nu}(\bm{q})$ has at least one flat band as the minimum energy band of the system, the SCGA predicts the stabilization of a classical spin liquid down to the lowest temperatures where the Lagrange multiplier $\lambda$ plateaus to the value corresponding to the fraction of low-energy flat bands times $n$~\cite{lozano_2024_atlas,lozano-2023}. For the classical spin liquid realized at the CTP discussed in the main text, the interaction matrix possesses four low-energy degenerate flat bands~\cite{lozano_2024_atlas}, which implies $\lambda\to 1$ in the low-temperature regime. It is worth noting that, to provide a more precise value for the separation between the intermediate and high temperature regimes, one may take the temperature $T^{\mathrm{MFT}}_{\mathrm{PM}}=\max_{m, \bm{q}}[\varepsilon_m(\bm{q})]/3$ as the mean-field temperature above which the paramagnetic phase is realized as discussed in Ref.~\cite{Enjalran_MFT_Pyrochlore}, where $\varepsilon_m(\bm{q})$ are the eigenvalues of the interaction matrix $\bm{J}^{\alpha\beta}_{\mu\nu}(\bm{q})$. Fig.~\ref{fig:lambda_T} illustrates the temperature evolution of the Lagrange multiplier $\lambda$ for three distinct parameter sets as indicated by the title of each panel. In these panels, the red dot marks the temperature at which the data shown in Figs.~\ref{fig:phasediagram-pinch-line}, \ref{fig:nematic_phase}, and \ref{fig:boundary_phase} are obtained.

\section{Details on calculating the pf-FRG phase diagrams}
\label{app:pffrg-details}

\begin{figure*}
    \centering
    \includegraphics{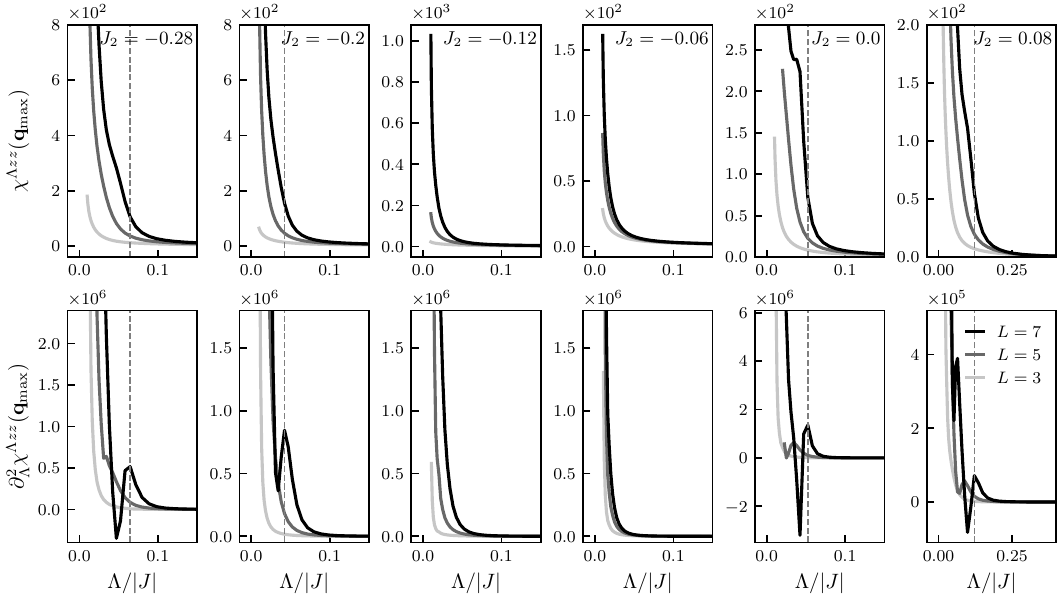}
    \caption{\textbf{RG flow of the spin-spin susceptibility} at the momentum $\bm{q}_\mathrm{max}$ where it is maximal for fixed $J_1/|J_3| = 0.12$, $J_4 = 0$ and $J_3 < 0$, but varying $J_2/|J_3|$. The dashed gray lines indicate the critical scale $\Lambda_c$ at which a flow breakdown is identified. The left and right most two panels are in the conventionally ordered $T_{1-}$ and $E$ phase, respectively and exhibit a putative flow breakdown manifesting as a nonmonotonicity in the second derivative of the flow. The center two plots show the flow in the nonconventional phase, which appears smooth and monotonous down to the lowest calculated $\Lambda/|J| = 0.01$.}
    \label{fig:susceptibility_flows}
\end{figure*}

\begin{figure}[h!]
    \centering
    \includegraphics{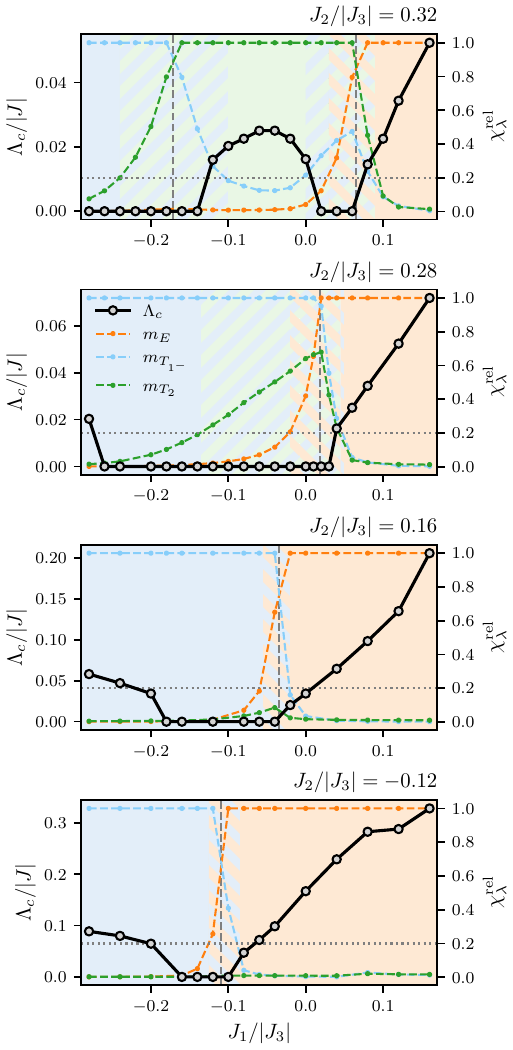}
    \caption{{\bf Cuts through the phase diagram in Fig.~\ref{fig:phasediagram-pinch-line}.} The black dots show the value of the critical scale $\Lambda_c$, the colored dots show the relative value of the irrep susceptibilities in the low-cutoff limit. Hatched background colors mark regions where multiple $\chi_\lambda^\mathrm{rel}$ are above 20\% (dotted horizontal line). The dashed lines mark the points where the largest irrep susceptibility changes.}
    \label{fig:phasediagram-cuts}
\end{figure}

\begin{figure*}
    \centering
    \includegraphics{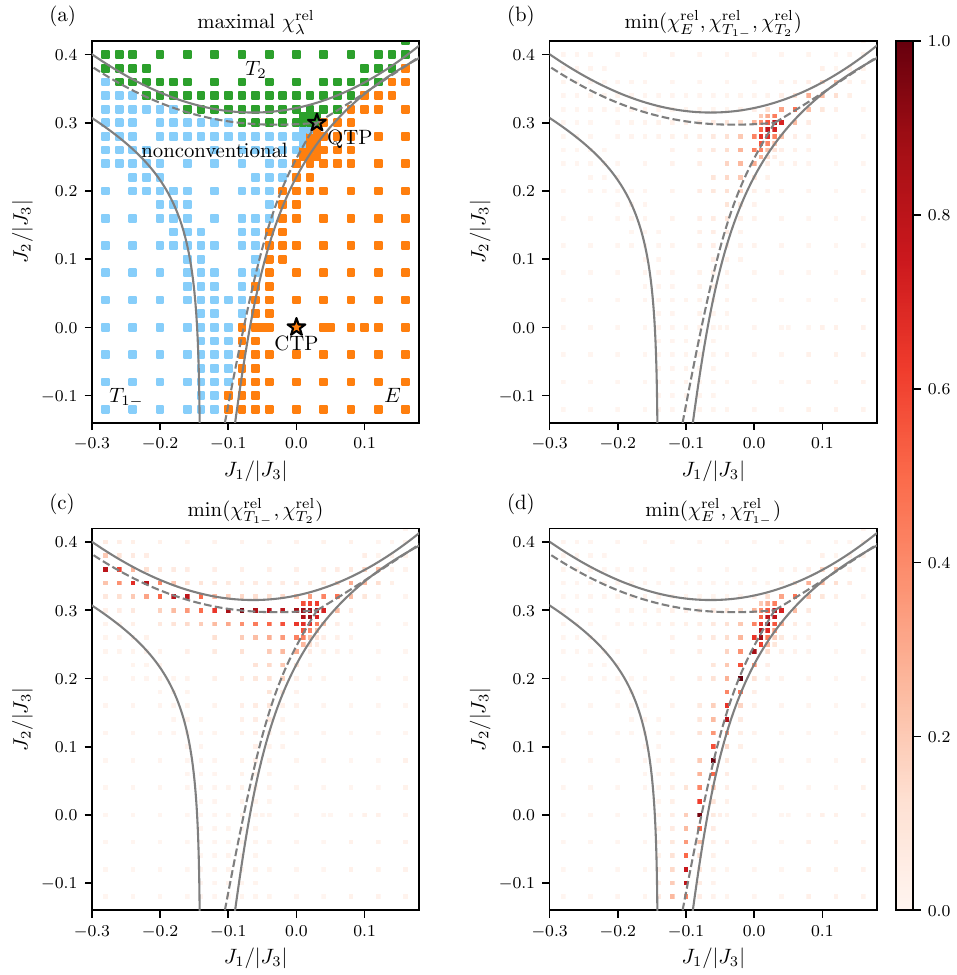}
    \caption{\textbf{Dominant order-parameter susceptibilities in the nonconventional regime}. (a) Maximal order-parameter susceptibility in the low-cutoff limit. (b)-(d) Minimal relative magnitude of the order-parameter susceptibility $\chi_\lambda^\mathrm{ref}$ [defined in Eq.~\eqref{eq:relative-irrep-susceptiblity}] between the orders characterizing the different nonconventional regimes (the spin nematic $T_{1-} \oplus T_2$, the pinch-line spin liquid $E\oplus T_{1-} \oplus T_2$ and the $E\oplus T_{1-}$ regime). The solid lines mark the approximate boundary of the nonconventional phase. The dashed lines highlight where the dominant order-parameter susceptibility changes and the degeneracy is maximal.}
    \label{fig:irrep_degeneracies}
\end{figure*}

\begin{figure}[t]
    \centering
    \includegraphics{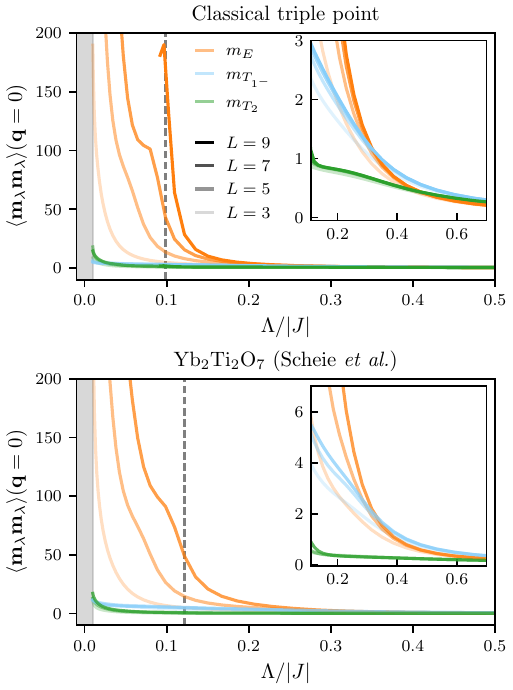}
    \caption{\textbf{Order-parameter susceptibility flows} at the CTP (top) and at the $\rm Yb_2Ti_2O_7$ parameters from Scheie \emph{et al.}~\cite{scheie2020} (bottom). The insets show the flows zoomed in at larger cutoffs. The clear dominance of $E$ order only emerges in the low-cutoff limit. The dashed gray lines indicate the critical scale $\Lambda_c$ at which a flow breakdown is identified.}
    \label{fig:irrep-flows}
\end{figure}

In this appendix, we give details on how the pf-FRG phase diagrams in Figs.~\ref{fig:phasediagram-pinch-line}(b) and \ref{fig:phase-diagram-ybti} were calculated from our pf-FRG data. 

\subsection{Discerning conventionally ordered from nonconventional phases}

Spontaneous symmetry breaking into dipolar order should, in theory, lead to a divergence of the relevant components of the RG susceptibility flow $\chi^{\Lambda \mu\nu}(\bm{q}_\mathrm{max})$ at a finite critical scale $\Lambda = \Lambda_c$ \cite{Muller-2024}. Here $\bm{q}_\mathrm{max}$ refers to the momentum $\bm{q}_\mathrm{max}$ where the susceptibility is maximal, characterizing the corresponding ground-state order. In our case, the diagonal components $\chi^{\Lambda \mu\mu}$ are always clearly dominant. Moreover, due to the symmetries of the pyrochlore lattice, they exhibit the same maximal flow (although at different symmetry-related momenta $\bm{q}_\mathrm{max}$). This allows us to restrict our flow breakdown analysis to the $zz$-component $\chi^{\Lambda zz}(\bm{q}_\mathrm{max}$). In practice, the numerical solution of the flow equations necessitates several approximations that often soften the expected divergence at $\Lambda_c$ to a kink or a hump. Additionally, these features tend to emerge only at sufficiently large system sizes, defined by the bond-length $L$, beyond which correlations are set to zero in the pf-FRG calculations. Hints of these features, however, can often be observed when analyzing not only the susceptibility flow itself, but also its second derivative $\partial^2_\Lambda\chi^{\Lambda\mu\nu}(\bm{q}_\mathrm{max})$. Initially small nonmonotonicities in the second derivative at small system sizes may grow to true divergencies for larger $L$. We therefore identify any nonmonotonicity in $\partial^2_\Lambda\chi^{\Lambda\mu\nu}(\bm{q}_\mathrm{max})$ that scales with increasing $L$ as a flow breakdown \cite{Gembe-2024}. As an example, the flows in Fig.~\ref{fig:susceptibility_flows} show the susceptibility flows and their second derivative along a horizontal cut through the phase diagram Fig.~\ref{fig:phasediagram-pinch-line}(b) for fixed $J_2/|J_3| = 0.12$ and Fig.~\ref{fig:phasediagram-cuts} shows the evolution of the critical scale for four additional cuts with fixed $J_2$. 

\subsection{Determining the dominant irrep susceptibilities}

In addition to distinguishing conventionally ordered from nonconventional phases, we characterize the nature of the ground-state spin-spin correlations by identifying the relevant order-parameter susceptibilities. The first step is to distinguish which order-parameter susceptibility is maximal in the low-cutoff limit ($\Lambda/|J| = 0.02$ provided the flow has not stopped at larger $\Lambda$ due to a flow breakdown), as shown in Fig.~\ref{fig:irrep_degeneracies}(a). This characterizes the dominant correlations in ground state. In the conventionally ordered phases, only a single order-parameter susceptibility clearly dominates (see, e.g. Fig.~\ref{fig:irrep-flows}). At the boundaries of the nonconventional phase, however, there appear small but extended regions where multiple order-parameter susceptibilities are relevant. This occurs in the putative nematic, $E\oplus T_{1-}$ and quantum pinch-line spin liquid regimes discussed in the main text. To gauge the extent of these regimes, we define relative susceptibilities by normalizing with respect to the maximum as
\begin{equation}
\label{eq:relative-irrep-susceptiblity}
    \left. \chi_{\lambda}^\mathrm{rel} = \frac{\langle \mathbf{m}_\lambda \cdot\mathbf{m}_\lambda\rangle}{\max_{\lambda^\prime} \langle \mathbf{m}_{\lambda^\prime}  \cdot\mathbf{m}_{\lambda^\prime}\rangle} \right|^{\Lambda\to0}_{\bm{q}=0}.
\end{equation}
In the phase diagrams shown in the main text, we highlight regions where more than one order-parameter susceptibility exceeds $\chi_\lambda^\mathrm{rel} > 0.2$ by a hatched background color and distinct markers. To illustrate this, Fig.~\ref{fig:phasediagram-cuts} shows the evolution of $\chi_\lambda^\mathrm{rel}$ as a function of $J_1$ for different fixed values of $J_2$. Note how the background colors change when any $\chi_\lambda^\mathrm{rel}$ crosses the 20\% threshold. To further highlight the extent of the regions where multiple order-parameter susceptibilities are degenerate, Fig.~\ref{fig:irrep_degeneracies}(b)-(d) show the minimal $\chi_\lambda^\mathrm{rel}$ relevant to the various nonconventional regimes. These show relatively clear lines of maximal degeneracy (shown with dashed gray lines) that occur at the boundary where the maximal order-parameter susceptibility changes, and meet at the quantum triple point.

We note that the dominant order-parameter susceptibility can change when considering larger RG cutoffs $\Lambda$ away from the low-cutoff limit. This is illustrated in Fig.~\ref{fig:irrep-flows}, which depicts the flow of the order-parameter susceptibilities at the CTP and the $\rm Yb_2Ti_2O_7$ parameters from Scheie \emph{et al.}~\cite{scheie2020}. In the classical model, the ground-state at the CTP exhibits degeneracy among $E$, $T_{1-}$ and $T_2$ order, while for $\rm Yb_2Ti_2O_7$ parameters the ground-state adopts $T_{1-}$ order but lies near the $E$ phase boundary (consistent with experimental findings for $\rm Yb_2Ti_2O_7$). In the quantum model, on the other hand, the ground state favors $E$ order in both cases. At large $\Lambda/|J| \approx 0.4$, however, the quantum and classical results are again consistent: at the CTP, all three susceptibilities are of similar magnitude, and for $\rm Yb_2Ti_2O_7$, $E$ and $T_{1-}$ dominate, with a slight advantage for $T_{1-}$. Only when integrating the flow equations to lower cutoffs -- and thereby incorporating more quantum fluctuations -- does the clear dominance of $E$ order emerge. This supports our argument made in the main text that the zero-temperature phase diagram may differ significantly from the finite temperature one, potentially explaining the discrepancies with experiments.

\section{Comparison with nonlinear spin-wave theory}
\label{app:nlswt-comparison}

\begin{table*}
    \begin{tabular}{c|c||c|c|c|c||c|c|c||c|c|c}
        Label & Reference (\emph{et al.}) & $J_1\, (\mathrm{meV})$ & $J_2\, (\mathrm{meV})$ & $J_3\, (\mathrm{meV})$ & $J_4\, (\mathrm{meV})$ & $J_1/|J_3|$ & $J_2/|J_3|$ & $J_4/|J_3|$ & $D/J$ & $(K + \Gamma)/J$ & $(K-\Gamma)/J$\\
        \hline
        (a) & Scheie \cite{scheie2020} & -0.026 & -0.307 & -0.323 & 0.028 & -0.08 & -0.95 & 0.087 & -0.631 & 0.014 & 0.151 \\
        (b) & Robert \cite{robert2015} &-0.03 & -0.32 & -0.28 & 0.02 & -0.107 & -1.143 & 0.071 & -0.68 & 0.03 & -0.039 \\
        (c) & Thompson \cite{thompson2017} &-0.028 & -0.326 & -0.272 & 0.049 & -0.103 & -1.199 & 0.18 & -0.525 & 0.11 & -0.096 \\
        (d) & Ross~\cite{ross2011} & -0.09 & -0.22 & -0.29 & 0.01 & -0.31 & -0.759 & 0.034 & -0.98 & 0.675 & 0.883 \\
    \end{tabular}
    \caption{\textbf{Exchange constants for Yb$_2$Ti$_2$O$_7$} in different parametrizations. The couplings were converted to the global frame ($J_1, J_2, J_3, J_4$) using Eq.~\eqref{eq:global-to-local} if not directly stated in the references, without accounting for uncertainties. The global parameters normalized by $|J_3|$ (with $J_3 < 0$) are the stars drawn in Fig.~\ref{fig:phase-diagram-ybti}. The dual global couplings parameterized by $\bar{D}, \bar{K}$ and $\bar{\Gamma}$ (normalized by $\bar{J} > 0$)  are the stars drawn in Fig.~\ref{fig:phase-diagram-rau}.}
    \label{tab:ybtio-couplings}
\end{table*}

One of our main observations -- the significant enlargement of the $E$-phase compared to the classical model -- is absent in a conventional linear spin-wave treatment \cite{yan2017}. However, a similar result was found using \emph{nonlinear} spin-wave theory (NLSWT) which includes the effects of magnon interactions \cite{rau19}.
Although for slightly different parameter sets compared to our calculations in the main text, they observed a breakdown of NLSWT in the $T_1$ phase near the classical boundary to the $E$ phase, resembling the shift seen in our pf-FRG calculations. To allow a direct comparison to their results (namely Fig.~7 in Ref.~\cite{rau19}), we have computed a pf-FRG phase diagram using exactly their parameters, which we describe in the following.

The calculations in Ref.~\cite{rau19} are performed in a different spin basis referred to as the \emph{dual} global frame [introduced before Eq.~\eqref{eq:dual-parameters}]. Instead of $\bar{J}_1, \bar{J}_2, \bar{J}_3, \bar{J}_4$, however, they use a parametrization of the exchange constants in terms of a dual Heisenberg $\bar{J}$, Kitaev $\bar{K}$, symmetric off-diagonal $\bar{\Gamma}$ and Dzyaloshinskii–Moriya $\bar{D}$ interaction defined as
\begin{equation}
        \bar{J} = \bar{J}_1,\quad
        \bar{K} = \bar{J}_2 - \bar{J}_1, \quad
        \bar{\Gamma} = \bar{J}_3, \quad
        \bar{D} = \sqrt{2} \bar{J}_4 \, .
\end{equation}
They compute a phase diagram for fixed $(\bar{K}-\bar{\Gamma})/J=-0.096$, matching the estimate for $\rm Yb_2Ti_2O_7$ from Thompson \emph{et al.} ~\cite{thompson2017} (see Table~\ref{tab:ybtio-couplings}). The classical boundary between the $T_1$ (in this frame $T_{1+}$) and $E$ phase then lies at $(\hat{K} + \hat{\Gamma})/\hat{J} = 0$, placing the parameters from Thompson \emph{et al.} well inside the classical $T_1$ phase. Their calculations, however, reveal a large region in the $T_1$ phase where NLSWT breaks down, extending from the classical $E-T_1$ boundary to just below the estimated $\rm Yb_2Ti_2O_7$ parameters. 

\begin{figure}[t]
    \centering
    \includegraphics{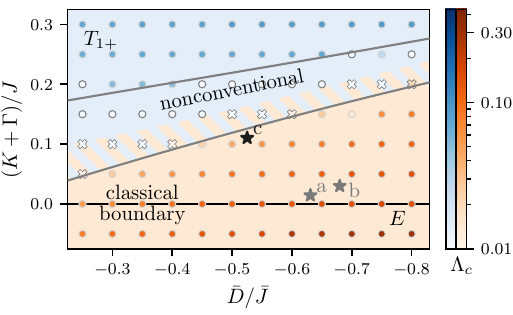}
    \caption{\textbf{Quantum phase diagram in correspondence with Ref.~\cite{rau19}} from pf-FRG. The coupling space is parameterized by the \emph{dual} parameters $\bar{J}, \bar{K}, \bar{\Gamma}, \bar{D}$ from Ref.~\cite{rau19}. The couplings $(\bar{K}-\bar{\Gamma})/\bar{J}_1=-0.096$ are fixed so that the parameters estimated for Yb$_2$Ti$_2$O$_7$ from Thompson \textit{et al.} \cite{thompson2017} lie in the plane. The stars show estimates for $(\bar{K}+\bar{\Gamma})$ and $\bar{D}$ from (a) Scheie \cite{scheie2020}, (b)  Robert \cite{robert2015} and (c) Thompson~\cite{thompson2017} \textit{et al.} (see Tab.~\ref{tab:ybtio-couplings}). The estimates from Ross~\cite{ross2011} \textit{et al.} lie far outside the shown parameter region.}
    \label{fig:phase-diagram-rau}
\end{figure}

Fig.~\ref{fig:phase-diagram-rau} shows the pf-FRG phase diagram for the same parameters. We again observe an enlarged $E$ phase and the emergence of a nonconventional phase between the $E$ and $T_1$ phase boundary. This phase contains ground-states with either dominant $T_{1+}$ correlations or mixed $E \oplus T_{1+}$ correlations. Its boundary closely resembles the boundary of the regime where NLSWT breaks down, but extends even further into the classical $T_1$ phase, placing $\rm Yb_2Ti_2O_7$ slightly in the $E$ phase. These findings are consistent with the discussion in Ref.~\cite{rau19}, in which the authors propose that the ground-state in the NLSWT unstable regime is possibly $E$ order, and that the extent of the unstable regime may be underestimated by their method. This agreement further supports their conclusion that magnon interactions beyond the linear-spin-wave treatment are crucial in driving the enlargement of the $E$-phase.

\section{Evolution of the low-energy Hamiltonian bands}
\label{sec:evolution_energy_bands}

\begin{figure*}
    \centering
    \begin{overpic}[width=1\textwidth]{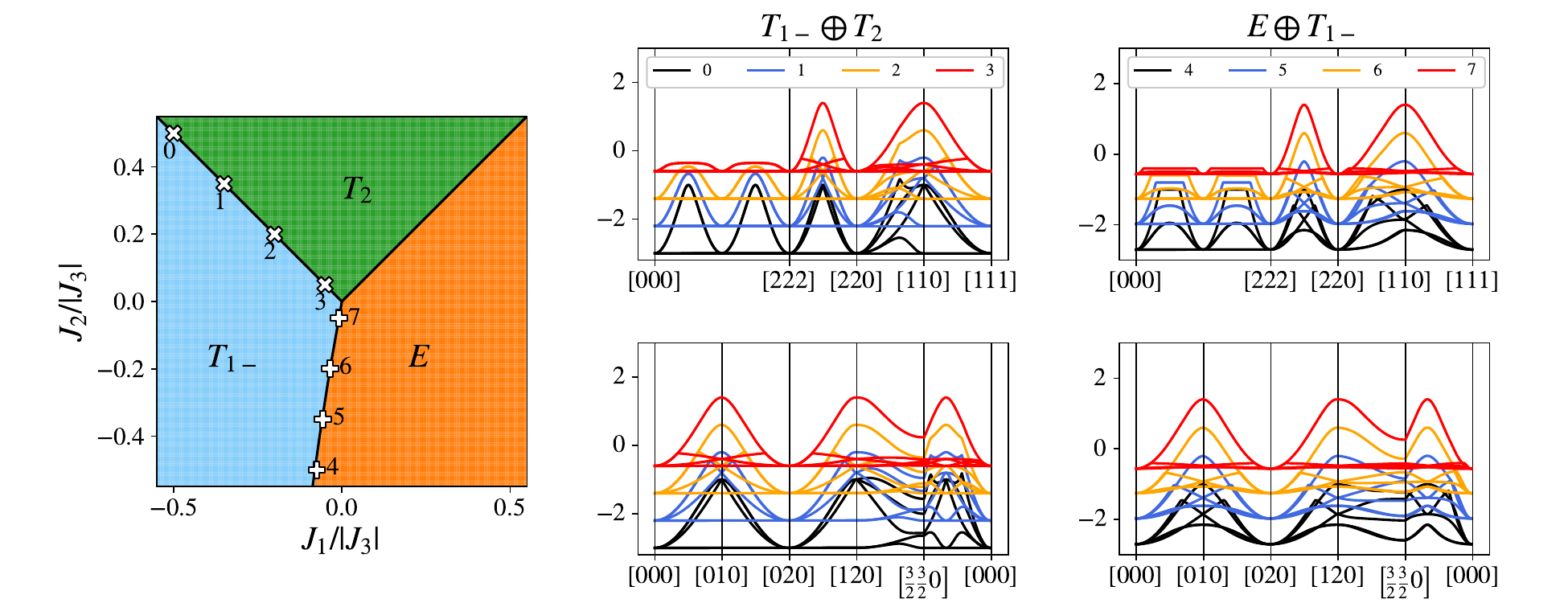}
     \put(9,32) {\textcolor{black}{(a)}}
     \put(39,36) {\textcolor{black}{(b)}}
     \put(39,17.5) {\textcolor{black}{(c)}}
     \put(70,36) {\textcolor{black}{(d)}}
     \put(70,17.5) {\textcolor{black}{(e)}}
    \end{overpic}
    \caption{{\bf Evolution of the lowest five energy bands of the Hamiltonian along (a) the $T_{1-}\oplus T_{2}$ line and (b) the $E\oplus T_{1-}$ line}. 
    Panel (a) shows the classical phase diagram of the bilinear spin model with $J_4=0$ and $J_3<0$.  
    The white `x' and `+' markers indicate four sets of parameters along the $T_{1-}\oplus T_{2}$ line and $E\oplus T_{1-}$ line, respectively.
    Panels (b) and (c) show the lowest five energy bands of the Hamiltonian in Eq.~\eqref{eq:hamiltonian}
    for a set of parameters along the $T_{1-}\oplus T_{2}$ line (left)  and the $E\oplus T_{1-}$ line (right). 
    The upper (lower) plot shows the evolution along a high-symmetry path in the $[hhl]$ ($[hk0]$) plane. In panels (b)-(e) the bands have been slightly shifted by multiples of $0.5$ from their actual low-energy value for ease of comparison. }
    \label{fig:bands_cuts}
\end{figure*}

In this appendix, we discuss the evolution of the low-energy bands of the interaction matrix $J_{ij}$ in Eq.~\eqref{eq:hamiltonian} along the $T_{1-}\oplus T_{2}$ and $E\oplus T_{1-}$  phase boundaries. We consider four distinct sets of parameters along these boundaries which progressively approach the classical triple point where the classical pinch-line spin liquid is stabilized; see Fig.~\ref{fig:bands_cuts}(a).

Fig.~\ref{fig:bands_cuts}(b) and (c) illustrate the five lowest energy bands of the interaction matrix $J_{ij}$ along the $T_{1-}\oplus T_{2}$ boundary for two sets of high-symmetry paths, one along the $[hhl]$ plane and another along the $[hk0]$ plane, respectively. The evolution of these bands identifies two prominent features: the observation of low-energy flat lines along high-symmetry directions (namely the $[111]$, $[010]$ directions, and symmetry-related directions), and the observation of band-touching points along these lines (observed at $[hkl]=[0,0,0]$ and $[hkl]=[1,1,1]$). We note that, as we progressively approach the classical triple point, the first four low-energy bands become flat, suggesting its proximity to a classical spin liquid.

Fig.~\ref{fig:bands_cuts}(d) and (e) illustrate the five lowest energy bands of the interaction matrix $J_{ij}$ along the $E\oplus T_{1-}$ boundary for two sets of high-symmetry paths, one along the $[hhl]$ plane and another along the $[hk0]$ plane. As was observed for the $T_{1-}\oplus T_{2}$ boundary, both flat lines and band-touching points along these lines can be observed. However, for the $E\oplus T_{1-}$ boundary, the flat lines are only observed along the $[111]$ direction with band-touching points with higher energy bands at  $[hkl]=[0,0,0]$ and $[hkl]=[1,1,1]$.

The observation of these low-energy features in the band spectrum has profound effects on the spin correlation functions predicted by the SCGA. Indeed, the correlation functions in this theory are proportional to the projection onto the low-energy modes of the $J_{ij}$ matrix, which, at low temperatures, result in stronger correlations at these points.

\FloatBarrier
\bibliography{pinch_line}

\end{document}